\documentstyle[eqsecnum,aps,prd]{revtex}

\begin{document}

\def\nbox#1#2{\vcenter{\hrule \hbox{\vrule height#2in
\kern#1in \vrule} \hrule}}
%e.g. \nbox{.1}{.1}
\def\sq{\,\raise.5pt\hbox{$\nbox{.09}{.09}$}\,}
\def\sqb{\,\raise.5pt\hbox{$\overline{\nbox{.09}{.09}}$}\,}

\title
{Weyl Cohomology and the Effective Action for Conformal Anomalies}
\author{Pawel O. Mazur\\Department of Physics and Astronomy,\\
University of South Carolina, Columbia SC 29208\\
{\em mazur@mail.psc.sc.edu}\\
\vspace{.15cm}
and\\
\vspace{.15cm}
Emil Mottola\\
T-8, Theoretical Division, MS B285\\
Los Alamos National Laboratory, Los Alamos, NM 87545\\
{\em emil@lanl.gov}\\
\vspace{.15cm}
LA-UR-01-3121}

\maketitle
\begin{abstract}

{\small{We present a general method of deriving the effective
action for conformal anomalies in any even dimension, which
satisfies the Wess-Zumino consistency condition by construction.
The method relies on defining the coboundary operator of the local
Weyl group, $g_{ab} \rightarrow \exp (2\sigma) g_{ab}$, and giving
a cohomological interpretation to counterterms in the effective
action in dimensional regularization with respect to this group.
Non-trivial cocycles of the Weyl group arise from local
functionals that are Weyl invariant in and only in the physical
even integer dimension $d=2k$. In the physical dimension the
non-trivial cocycles generate covariant non-local action
functionals characterized by sensitivity to global Weyl
rescalings. The non-local action so obtained is unique up to the
addition of trivial cocycles and Weyl invariant terms, both of
which are insensitive to global Weyl rescalings.  These distinct
behaviors under rigid dilations can be used to distinguish between
infrared relevant and irrelevant operators in a generally
covariant manner. Variation of the $d=4$ non-local effective
action yields two new conserved geometric stress tensors with
local traces equal to the square of the Weyl tensor and the
Gauss-Bonnet-Euler density respectively. The second of these
conserved tensors becomes $^{(3)}H_{ab}$ in conformally flat
spaces, exposing the previously unsuspected origin of this tensor.
The method may be extended to any even dimension by making use of
the general construction of conformal invariants given by
Fefferman and Graham. As a corollary, conformal field theory behavior of 
correlators at the asymptotic infinity of either anti-de Sitter or de Sitter 
spacetimes follows, {\it i.e.} AdS$_{d+1}$ or deS$_{d+1}$/CFT$_d$ correspondence. 
The same construction naturally selects all infrared relevant terms (and only those 
terms) in the low energy effective action of gravity in 
any even integer dimension. The infrared relevant terms arising from the known 
anomalies in $d=4$ imply that the classical Einstein theory is modified at large
distances.}}

\end{abstract}
\section{Introduction}

The existence of conformal or trace anomalies in quantum field theories has
been known for several decades \cite{genanom,DDI,duffa,Bir}. Nevertheless, their
full mathematical structure as well as their physical implications have
remained a subject of discussion \cite{duffb}. Accordingly, our purpose in this
paper will be twofold. First, we wish to clarify the mathematical structure of
conformal anomalies as non-trivial cocycles of the cohomology of the Weyl
group, by defining the Weyl coboundary operator acting on functionals of the
spacetime metric, and its corresponding cochain. This definition of the
cohomology of the Weyl group will enable us to show that the trace anomaly
terms are in one-to-one correspondence to local counterterms in the
dimensionally regulated effective action in $d$ dimensions, which become Weyl
invariant in {\it and only in} the physical even dimension $d=2k$. The
limit, $d\rightarrow 2k$ gives rise to a well defined non-local
action for each non-trivial cocycle in the physical dimension, which is unique
up to Weyl invariant and cohomologically trivial terms. In addition to
supplying a well-defined and efficient algorithm for calculating the effective
action of trace anomalies in any even dimension (provided that the local Weyl
invariants in that dimension are known), the use of dimensional regularization
will serve our second purpose, which is to expose the physical implications of
the resulting anomalous effective action. In fact, exactly the same property of
the effective action that identifies it as a non-trivial cocycle of the Weyl
group, its multi-valuedness, also implies that it is sensitive to global Weyl
rescalings, and therefore infrared relevant in the Wilson renormalization
group sense. That is, the elements of the non-trivial cohomology of the Weyl
group necessarily give rise to terms in the low energy effective action of
gravity which do not decouple in limit of low energies or large
spacetime volumes. Specializing to $d=4$ spacetime dimensions, this
means that the known trace anomalies necessarily imply modifications
of classical general relativity in the effective low energy, long
distance theory of gravity.

The relationship of anomalies to conformal invariants in even integer spacetime
dimensions allows for a broader connection between pure mathematics and the
physics of the renormalization group and low energy effective actions, implicit
in the recent literature on AdS/CFT correspondence \cite{adscft}. Fefferman and Graham
(hereafter referred to as FG) provided an algorithm for constructing conformal
invariants in $d$ dimensions by embedding the physical space of interest in a
$d+2$ dimensional Ricci flat ambient space, in such as way that coordinate
invariant scalars in the embedding space (which are easy to construct) give
rise to conformal Weyl invariants (which are generally more difficult to find)
in the physical space \cite{FG}. The construction makes use of the conformal
properties of the light cone in the $d+2$ dimensional ambient space and the
foliation of this space by $d+1$ dimensional surfaces of asymptotically
constant Ricci curvature which approach this light cone in a specific
coordinatization. The FG algorithm is exactly what is needed to generate local
conformal invariants and hence non-trivial cocycles of the Weyl group in any
even dimension in accordance with our general dimensional regularization
method.

Furthermore, because the physical space of interest lies at a conformal
boundary of the $d+1$ dimensional bulk or embedding space, conformal
transformations of the boundary form a special class of coordinate
transformations in the bulk space of asymptotically constant curvature. The
geometric radial coordinate of the bulk space geometry $\rho$ corresponds to the
length scale of the finite size rescalings in the effective theory on the
boundary in a precise way. The set of local diffeomorphism invariants in the
bulk action which diverge as the conformal boundary is approached generate
exactly the infrared relevant operators of the Wilson effective action on the
boundary, local or not, and a shift in the $\rho$ coordinate of the bulk
geometry is precisely a finite size scale transformation in the $d$ dimensional
boundary theory. The infrared divergent terms as $\rho \rightarrow 0$ may also
be regularized by dimensional continuation. Thus the FG embedding together with
dimensional regularization provide just the mathematical tools necessary to define the
infrared relevant operators of the Wilson effective action in a theory with general
coordinate invariance in a precise way.

A by-product of this study of conformal infinity in the FG coordinates is
the conclusion that the structure of conformal field theories in the AdS/CFT
correspondence is actually representative of a more general feature of the FG
embedding, which requires neither AdS (since it works equally well for
asymptotically de Sitter spaces), nor a supersymmetric CFT on the conformal
boundary. The conformal behavior at infinity (as opposed to specific values
of coefficients in the effective action or correlation functions) and
the generation of infrared relevant terms in the effective
gravitational theory asymptotically as $\rho \rightarrow 0$
is a purely kinematic property of the FG embedding of physical spacetime in a
space of one higher dimension with constant positive or negative
scalar curvature. Hence, there is a deS/CFT correspondence as well
as an AdS/CFT correspondence, although the former has no evident
connection with string theory backgrounds, as in the AdS/CFT case \cite{adscft,Mal}.

Before presenting the general method and exploring these relationships in
detail, let us review some well known features of conformal anomalies and
previous treatments of their cohomology. The standard route to exhibiting
the algebraic structure of the trace anomaly is to treat the parameters
of infinitesimal local conformal transformations as anti-commuting Grassmann
variables \cite{Bon,Breg}. This guarantees nilpotence of the generator of the
infinitesimal  Weyl transformations, and distinguishes trivial from
non-trivial cocycles in a  clear manner. However, an important aspect of the
anomaly is left unexplored in this abstract algebraic approach. Non-trivial
cohomology implies that although the conformal anomaly itself is local, it
cannot be written as the Weyl variation of some local coordinate invariant
action. Instead the action whose Weyl variation is the local anomaly must
itself be non-local in any even integer dimension. The Wess-Zumino (WZ)
consistency condition is just the integrability condition that this coordinate
invariant non-local action exists \cite{WZ,AMMnp}. Consistency is automatically
satisfied in the algebraic approach by the nilpotence of two successive
anti-commuting Weyl transformations, but the algebraic method furnishes no
means of constructing the WZ consistent effective actions associated with the
non-trivial cocycles, and sheds little light on their physical meaning.

A complementary approach to conformal anomalies is to construct the WZ effective
action directly by `integrating the trace anomaly' with respect to the local
conformal factor variation $\delta \sigma (x)$. In two dimensions this
procedure is almost immediate and can be performed by inspection \cite{Poly},
while in four and higher even integer dimensions it requires adding
a admixture of the trivial cocycles ({\it i.e.} those terms which {\it can} be
written as the Weyl variation of a local coordinate invariant action) in order
to bring the anomaly into a form linear in $\sigma (x)$, which can be
integrated easily \cite{Rieg}.  Although this approach certainly produces an
action satisfying WZ consistency \cite{AMMnp}, it leaves the deeper cohomological
aspects of conformal anomalies unexamined. Moreover, since it produces a WZ action
that involves an $d^{\rm th}$ order differential operator on $\sigma$ in $d=2k$ even
dimensions, it seems to imply strong ultraviolet (UV) behavior, and raises concerns about
ghosts similar to local higher derivative theories of gravity in dimensions greater than
two. Conversely and perhaps paradoxically, the non-local form of the WZ action obtained
in this approach seems to imply more severe IR behavior than expected with unphysical
$p^{-4}$ poles in anomalous Ward identities \cite{Desera}. Yet the anomaly is the effect
of integrating out massless fields in all the standard one-loop calculations and the fully
covariant non-local form of the WZ anomalous action should be associated with
long distance or low energy physics, in which there can be neither unphysical UV ghosts
nor higher order IR poles. Hence this approach raises a number of questions which should
be resolved before it can be deemed completely satisfactory.

In this paper we present a general, unified treatment of conformal anomalies
in any even dimension by using the same method both to define the non-trivial
cohomology of the Weyl group and to derive the local WZ effective action, as
well as its fully covariant non-local form. The key observation which will
make this possible is that the non-trivial cocycles of the Weyl cohomology in
any even integer dimensions are in one-to-one correspondence with those local
counterterms in the dimensionally regulated effective action which become
Weyl invariant in {\it and only in} exactly the physical dimension $d=2k$.

The technical reason for this one-to-one correspondence is that such local
invariants have Weyl variations which are proportional to $d-2k$ near the
physical dimension and hence cancel the $(d-2k)^{-1}$ pole of dimensional
regularization, yielding a {\it finite} action $\Gamma_{WZ}$ for $d=2k$, which
is local in terms of $\sigma$, and which automatically satisfies the WZ
consistency condition. This becomes clear when $\sigma$ is eliminated from the
action by solving for the unique conformal factor that Weyl translates
between two different metrics $\bar g_{ab}$ and  $g_{ab} \equiv
e^{2\sigma}\bar g_{ab}$ in the same conformal equivalency class.  Then the
local WZ action becomes the difference of two fully covariant but non-local
actions,
\begin{equation}
\Gamma_{WZ}[\bar g; \sigma] = S_{anom}[\bar g e^{2\sigma}] - S_{anom}[\bar g]\,.
\end{equation}
This difference may be viewed as a finite Weyl coboundary operation,
$\Delta_{\sigma} \circ S_{anom}[\bar g]$, which produces a one-form from
a scalar functional $S$ \cite{fadsha}. The anti-symmetrized coboundary operator
on  $k$-forms may be defined by a straightforward generalization from standard
Riemannian geometry, and the nilpotency condition of $\Delta_{\sigma}$ on the
cochain easily checked ({\it cf.} Sec. 2). Hence the WZ consistency of
$\Gamma_{WZ}$ follows immediately. Its cohomology is nevertheless non-trivial
since $S_{anom}[g]$ is a non-local functional of $g_{ab}$  when written
entirely in the physical dimension $d=2k$, and $\Gamma_{WZ} [\bar g; \sigma]$
is not a single-valued functional of the original metric, $\bar g_{ab}$. This
multi-valuedness of the configuration space of metrics is due to the exclusion
of the singular metrics $g_{ab} = 0$ and its inverse, $g^{ab} =0$, corresponding to
vanishing or diverging conformal factor $\Omega = e^{\sigma}$. These singular
metrics correspond to punctures that allow the topological Euler number of the
manifold to change. Because of this topological obstruction in the
configuration space of smooth metrics, $\Gamma_{WZ} [\bar g; \sigma]$ is not
invariant under the shift $\sigma(x) \rightarrow \sigma (x) + i\pi q$,
corresponding to integer $q$ winding around the obstruction, and for the same
reason $\Gamma_{WZ} [\bar g; \sigma]$ is necessarily non-invariant under the
constant global scale transformation, $\sigma (x) \rightarrow \sigma (x) +
\sigma_0$. It is this sensitivity to finite volume scaling of
the effective WZ action which reveals its physical interpretation in
terms of standard Wilson renormalization group principles \cite{Wil}.

The use of dimensional regularization to define the cohomology of
the Weyl group has a number of advantages. From a purely mathematical
point of view, it reduces the general classification of the non-trivial
cocycles, usually associated with quantum anomalies, to the
construction of conformal invariants, a problem of classical
differential geometry that has been (nearly) rigorously solved in any even
dimension by Fefferman and Graham \cite{FG}. Since the non-trivial cocycles
and the corresponding non-local effective action $S_{anom}$ are {\it unique}
up to addition of trivial cocycles and completely Weyl invariant terms, the
explicit construction of the WZ effective action eliminates doubts
about the correctness of the procedure of integrating the anomaly. As a
consequence, in $d=4$ we can conclude definitively there are {\it no} non-local
terms in the effective action, of the form $C_{abcd}\log (-\sq)C^{abcd}$, generated
by the anomaly, as has long been conjectured \cite{DDI,Desera,DesSch}. Moreover,
dimensional regularization is the natural way to regularize the solutions of the
Einstein equations on embedding spaces of asymptotically constant curvature, as
required by the FG algorithm for constructing conformal invariants. Hence
continuation in the number of dimensions removes the `FG ambiguity,' regulates
the infrared divergences at large volumes in the FG construction, and also
illuminates why the trace anomaly of conformal field theories is given
by the AdS/CFT correspondence conjecture \cite{adscft,HenSken}.

{}From a more physical point of view, dimensional regularization is closely connected
to renormalization and the renormalization group (RG) flow of a quantum theory,
and allows a treatment of UV and IR renormalization effects in a unified way.
It is well known in statistical physics and renormalization theory that UV behavior
is enhanced by increasing the number of spacetime dimensions in loop integrations
while conversely, the IR behavior is enhanced by decreasing $d$ in a
given graph. Since $d$ is not required {\it a priori} to be greater than or
less than the physical dimension $2k$ in the dimensional regularization
procedure, the limit $d\rightarrow 2k$ of counterterms in the quantum effective action
can (and does) contain both UV and IR logarithmic effects in the physical dimension,
treated on an equal footing. In the quantum effective action poles replace
the UV cutoff and IR effects appear as finite terms, with no singular behavior
as $d$ approaches the physical dimension. Conversely, from the Wilson infrared effective
action point of view, the UV cutoff is fixed and terms in the effective action
are classified according to their behavior under rigid scale transformations
with a sliding IR cutoff, which we may take to be the total $d$-volume of
the system. Dimensional continuation may be used again, this time to regulate the
IR behavior, preserving generally coordinate invariance as the volume is taken to
infinity. In this case, poles signal large volume divergences or IR relevant
operators in the Wilson RG approach, while terms which remain finite as the volume
goes to infinity are IR irrelevant and contain no poles at the physical dimension.

These considerations and the non-single valuedness of $\Gamma_{WZ}$ under global scale
transformations imply that the non-trivial Weyl cocycles are best understood as infrared
marginally relevant operators of the Wilson effective action of low energy gravity under
the RG, which control the critical exponents and scaling behavior of the theory near its
IR conformal fixed point(s). These fixed point(s) are Gaussian because of the quadratic
form of  $\Gamma_{WZ}$, provided that the coefficients of the trivial cocycles
({\it i.e.} the UV relevant counterterms) flow to zero in the extreme
infrared \cite{AMa,AO}.

This physical RG interpretation of conformal anomalies is considerably more general
than existing anomaly calculations in classically conformally invariant free
field theories, and implies that the appearance of $S_{anom}$ in the
effective action of low energy gravity with {\it some} coefficient is quite
generic within the standard framework of low energy effective field theories.
The RG scaling interpretation is completely consistent with the scaling
behavior of $d=2$ CFT's coupled to gravity both in the continuum and in
lattice simulations, where the KPZ critical exponents are determined by the
Polyakov action \cite{KPZ,DTs}. In $d=4$ it is consistent with the IR stability of
the Gaussian fixed point \cite{AMa}, with the recent consideration of the
scaling behavior of the effective action of a massless, but not conformally
coupled scalar field in de Sitter space \cite{And}, and with the fact there are
{\it no} higher order poles or unphysical propagating ghost states in the $d=4$
effective quantum theory determined by $\Gamma_{WZ}$ \cite{states}.

A final bonus of the complete classification of the Weyl cohomology by
dimensional regularization in any even dimension is that it automatically
yields the conserved energy-momentum tensor corresponding to the non-local
action $S_{anom}[g]$. Although these tensors are also non-local in general,
in four dimensions one of them becomes the local geometric tensor
$^{(3)}H_{ab}$, defined by (\ref{Hthree}) in $d=4$ conformally flat spacetimes,
which had been found some time ago \cite{GinKir}. Thus the existence of
the covariant non-local action functional $S_{anom}$ turns out to be the underlying
reason for the existence of this local tensor, which has been called `accidentally
conserved' \cite{Bir}. Its non-local generalization to non-conformally flat
spacetimes agrees with the direct calculation from $\Gamma_{WZ}$ given in ref.
\cite{states}, and its trace is proportional to the Gauss-Bonnet-Euler density in
$d=4$.

The paper is organized as follows. In the next section we define the
anti-symmetric coboundary operator of finite Weyl shifts, generalizing
the ideas of differential forms and de Rham cohomology to the functional
space of metrics, without the use of infinitesimal anticommuting Grassmann
numbers. In Section 3 we illustrate our general technique for constructing
$\Gamma_{WZ}$ in $d=2$ dimensions, recovering both the Polyakov action
and the energy-momentum tensor corresponding to it. In Section 4 we follow
the same procedure in $d=4$ dimensions, deriving the form of $\Gamma_{WZ}$
previously found by integrating the anomaly, as well as its fully covariant
non-local form, $S_{anom}$. The conserved geometrical stress tensors
corresponding to this action are derived in Section 5, where the connection to
the tensor $^{(3)}H_{ab}$ is also demonstrated. Section 6 contains a brief
mathematical interlude, recapitulating the FG embedding in spaces of
asymptotically constant curvature, the conformal infinity of both AdS and
de Sitter space in suitable coordinates, and the subset of diffeomorphisms
in the bulk or embedding space which become local Weyl transformations on the
conformal boundary. In Section 7 the FG construction is applied to finite volume scale
transformations, which serves to select exactly the infrared relevant operators of the
effective Wilson gravitational boundary action. Section 8 summarizes our conclusions
and contains the main results of the paper in a concise form.

Wherever feasible detailed formulae used in the main text are relegated to the
three Appendices. The first catalogs the conformal variations of various tensors
needed in the text, the second contains a proof of an interesting
identity of the Weyl tensor in four dimensions, and the third is a computation
of the new non-local tensor $C_{ab}$ that appears in the variation of the
$d=4$ effective action in the general case.

\section{Cohomology of the Weyl Group and Dimensional Regularization}

We consider the abelian Weyl group of local conformal transformations on the metric,
\begin{equation}
\bar g_{ab}(x)\ \rightarrow\ g_{ab}(x) \equiv e^{2\sigma (x)}\ \bar g_{ab}(x)\,,
\label{weyl}
\end{equation}
where $\sigma (x)$ is any smooth function of the coordinates. Let $S[g]$ denote any
scalar functional (local or not) of the metric $g_{ab}$. The action of the Weyl group
(\ref{weyl}) suggests that a finite difference or coboundary operator $\Delta_{\sigma}$
on scalar functionals $S[g]$ be defined by
\begin{equation}
\Delta_{\sigma}\circ S[\bar g] \equiv (\Delta S)_{\sigma} \equiv S[e^{2\sigma} \bar g]
- S[\bar g]\,.
\label{zeroweyl}
\end{equation}
By definition, functionals $S_{inv}$ which are invariant under the Weyl
transformation (\ref{weyl}) satisfy $\Delta_{\sigma} \circ S_{inv} = 0$.

In general, $\Delta_{\sigma} \circ S \neq 0$, and this quantity depends on
both the initial or fiducial metric, $\bar g_{ab}(x)$ and the finite Weyl shift
$\sigma (x)$. It may be regarded as a one-form, $\Gamma^{(1)}[\bar g;\sigma]$
with respect to the finite shift Weyl group \cite{fadsha}. This is a natural
generalization of the concept of a one-form from the differential geometry of
finite dimensional Riemannian manifolds to the present case of the infinite
dimensional functional  space of metrics. The definition of the coboundary
operator $\Delta_{\sigma}$ above acting on scalar functionals
can be generalized to  $\Delta_{\sigma_{k+1}}$ acting on $k$-forms
$\Gamma^{(k)}[\bar g; \sigma_1, \dots, \sigma_k]$ by the anti-symmetrized chain rule,
\begin{equation}
\Delta_{\sigma_{k+1}}\circ \Gamma^{(k)}[\bar g; \sigma_1,\dots ,\sigma_k] \equiv
\sum_{i=1}^{k+1} (-)^{i+1} (\Delta\Gamma^{(k)})_{\sigma_i}[\bar g; \sigma_1,\dots ,
\sigma_{i-1}, \sigma_{i+1}, \dots,\sigma_{k + 1}]\,,
\label{kweyl}
\end{equation}
where
\begin{eqnarray}
(\Delta\Gamma^{(k)})_{\sigma_i}[\bar g; \sigma_1,\dots ,\sigma_{i-1},
\sigma_{i+1}, \dots,\sigma_{k+1}] &\equiv& \Gamma^{(k)}[\bar ge^{2\sigma_i}; \sigma_1,
\dots ,\sigma_{i-1}, \sigma_{i+1}, \dots,\sigma_{k+1}]
\nonumber\\
&&\qquad - \Gamma^{(k)}[\bar g; \sigma_1,
\dots ,\sigma_{i-1}, \sigma_{i+1}, \dots,\sigma_{k+1}]\,,
\label{finweyl}
\end{eqnarray}
is the Weyl finite difference operator on the functional,
$\Gamma^{(k)}[\bar g; \sigma_1,\dots ,\sigma_{i-1}, \sigma_{i+1}, \dots,\sigma_{k+1}]$
with $\sigma_i$ omitted, treated as a scalar functional of the metric $\bar g$ with all
the other $\sigma$ dependences unaffected. This anti-symmetrized
coboundary operation (\ref{kweyl}) produces a $(k+1)$-form from a $k$-form,
which is totally anti-symmetric under interchange of any two of its arguments,
$\sigma_i$ and $\sigma_j$, and defines a finite derivation operation in the
space of metrics with the same algebraic properties as the exterior derivative
in Riemannian geometry. If the Weyl transformation by $\sigma$, (\ref{weyl}) is
replaced by an infinitesimal anticommuting Grassmann variable, then the
definition (\ref{kweyl}) becomes equivalent to the standard definition \cite{Bon,Breg}.
The anti-commuting Grassmann variables are simply a device to keep track of
anti-symmetrization, and are not essential, provided the coboundary
operator $\Delta_{\sigma}$ is defined in an explicitly anti-symmetric manner
for a {\it finite} $\sigma$ Weyl shift, as in (\ref{kweyl}). The previous definition
(\ref{zeroweyl}) is a special case of the general definition (\ref{kweyl}) with $k=0$,
or in other words, zero-forms $\Gamma^{(0)} [\bar g]$ are just coordinate invariant
scalar functionals of the metric. The  Einstein-Hilbert action of classical general
relativity or the general coordinate  invariant effective action of a quantum
field theory $S_{eff}[\bar g]$ in a general curved  background metric $\bar g$
are examples of such coordinate invariant scalar functionals of the metric.

The definition of the cochain (\ref{kweyl}) allows us to give a precise meaning
to the cohomology and WZ consistency for the Weyl group. Applying the general
definition for $k=1$, the operation of the coboundary operator 
$\Delta_{\sigma}$ on one-forms produces the two-form,
\begin{eqnarray}
\Gamma^{(2)}[\bar g; \sigma_1,
\sigma_2]&=&\Delta_{\sigma_2}\circ \Gamma^{(1)}[\bar g; \sigma_1] =
(\Delta\Gamma^{(1)})_{\sigma_1}[\bar g;\sigma_2] -
(\Delta\Gamma^{(1)})_{\sigma_2}[\bar g;\sigma_1]\nonumber\\ &=&
\Gamma^{(1)}[\bar ge^{2\sigma_1}; \sigma_2] -\Gamma^{(1)}[\bar g; \sigma_2]  -
\Gamma^{(1)}[\bar ge^{2\sigma_2}; \sigma_1] + \Gamma^{(1)}[\bar g;
\sigma_1]\,. \label{twoweyl} \end{eqnarray} If $\Gamma^{(1)}[\bar g;
\sigma_1]$ is the one-form (\ref{zeroweyl}), {\it i.e.} if we apply
(\ref{twoweyl}) to the case, $\Gamma^{(1)}[\bar g;\sigma_1] =
\Delta_{\sigma_1}\circ S[\bar g]$, then we obtain the algebraic identity,
\begin{equation} (\Delta_{\sigma_2}\circ \Delta_{\sigma_1})\circ S[\bar g]
\equiv \Delta_{\sigma_2}\circ (\Delta_{\sigma_1}\circ S[g]) = 0\,.
\label{nilzero} \end{equation} This nilpotency property of $\Delta_{\sigma}$
is easily seen to be generally true on any $k-$form, owing to the
anti-symmetry of the general definition (\ref{kweyl}). That is, quite
generally we have the nilpotency property,
\begin{equation}
(\Delta^2)_{\sigma_1,\sigma_2} \equiv \Delta_{\sigma_1}\circ \Delta_{\sigma_2} = 0\,,
\label{nilp}
\end{equation}
an essential property of a coboundary operator, which justifies the use
of this term from ordinary differential geometry.
To define the cohomology, let us call any one-form $\Gamma^{(1)}[\bar g;\sigma_1]$
which satisfies
\begin{equation}
\Delta_{\sigma_2}\circ \Gamma^{(1)}[\bar g;\sigma_1] = 0\qquad {\rm closed},
\label{close}
\end{equation}
and any one-form $\Gamma^{(1)}[\bar g;\sigma_1]$ that can be written,
\begin{equation}
\Gamma^{(1)}[\bar g;\sigma_1] = \Delta_{\sigma_1}\circ S_{local}[\bar g]\qquad {\rm
exact},
\label{exact}
\end{equation}
if $S_{local}[g]$ is some local, single-valued scalar functional of the metric.
Eq. (\ref{nilzero}) shows that all exact one-forms are closed, because of the
nilpotency property (\ref{nilp}) of the coboundary shift operator $\Delta_{\sigma}$.
This is the trivial cohomology of the Weyl group. However, all closed forms are not
necessarily exact, and the non-trivial cohomology of the Weyl group is defined precisely
as the set of these closed but {\it non}-exact one-forms. That is, elements of the
non-trivial cohomology of the Weyl group are one-forms obeying (\ref{close}) that
{\it cannot} be written as Weyl transforms of any local, single-valued
functional of the metric, as in (\ref{exact}).

The insistence on local single-valued functionals of the metric, $S_{local}[g]$
for the exactness property (\ref{exact}) is essential. The underlying reason is
that the measure of integration over the space of metrics is single-valued and
local, so that any $S_{local}[g]$ can be absorbed into the definition of the
functional measure, whereas non-local functionals in the effective action
cannot be so absorbed, and constitute the genuine anomalies. The Weyl group
involves the exponential map (\ref{weyl}), which is certainly single-valued
and local. However, the inverse of the exponential map is a logarithm, which
is not single-valued, and as we shall see, non-locality results from solving
for $\sigma$ in terms of the original metric $\bar g_{ab}$ and its Weyl translate
$g_{ab}$, which involves inverting a differential operator, a non-local operation.
Hence, arbitrary functionals $\Gamma^{(1)}[\bar g;\sigma_1]$ of $\bar g$ and
$\sigma$ separately (even if restricted  to purely local ones) need not be
single-valued, local functions of the full Weyl transformed metric
$g= e^{2\sigma}\bar g$. These are the closed but non-exact functionals which
determine the non-trivial cohomology of the Weyl group, and for which we seek
an algorithm to compute explicitly.

As the non-exactness of one-forms defines the non-trivial cohomology of the
Weyl group and corresponds to genuine conformal anomalies, the condition of
closure (\ref{close}) corresponds precisely to the WZ consistency condition on
the anomalous effective action. Indeed let $\Gamma_{WZ}[\bar g;\sigma]$ be the
one-form effective action whose variation generates the Weyl anomaly $\cal A$
in the full (Weyl-transformed) metric $g_{ab}= \bar g_{ab} e^{2\sigma}$, {\it
i.e.}
\begin{equation}
{\delta \over \delta\sigma}
\Gamma_{WZ}[\bar g;\sigma] = {\cal A}[\bar g e^{2\sigma}]\,.
\label{AWZ}
\end{equation}
Then the statement that $\Gamma_{WZ}[\bar g;\sigma]$ is closed,
\begin{eqnarray}
\Delta_{\sigma_2} \circ \Gamma_{WZ}[\bar g;\sigma_1] &=&
\Gamma_{WZ}[\bar ge^{2\sigma_1};\sigma_2] -
\Gamma_{WZ}[\bar g;\sigma_2] -
\Gamma_{WZ}[\bar ge^{2\sigma_2};\sigma_1] +
\Gamma_{WZ}[\bar g;\sigma_1]\nonumber\\
&=& 0\,,
\label{WZ}
\end{eqnarray}
is precisely the statement that $\Gamma_{WZ}[\bar g;\sigma]$ satisfies
the WZ consistent anomaly condition \cite{WZ,AMMnp}.
It is the finite shift generalization of the infinitesimal form of WZ
integrability condition for the abelian Weyl group, {\it i.e.}
\begin{equation}
{\delta^2 S\over \delta\sigma_2\delta\sigma_1} =
{\delta^2 S\over \delta\sigma_1\delta\sigma_2}\,.
\label{infwz}
\end{equation}
Although this is guaranteed algebraically by the use of anti-commuting infinitesimal
Weyl parameters, it is important to recognize that the physical content
of the condition (\ref{infwz}) is that some scalar effective action functional
$S[\bar g]$ must exist, whose first variation is the anomaly, and whose
second variation is necessarily hermitian with respect to interchange
of the order of successive infinitesimal Weyl variations \cite{WZ,AMMnp}.

With these preliminaries we come now to the essential observation
which will enable us to give a general technique for constructing the
WZ consistent anomalous effective actions in any even $d=2k$ spacetime
dimension, by using well-known properties of dimensional regularization. In
dimensional regularization we are instructed to write down all the local
curvature invariants in $d$ dimensions near the physical $d=2k$ dimension,
multiplied by simple poles in $d-2k$, as allowed  counterterms for the
effective gravitational action of a quantum theory. These local $d$
dimensional counterterms are not Weyl invariant in general. If they are
not Weyl invariant then their Weyl derivation by $\Delta_{\sigma}$ will
still contain a pole at $d=2k$, indicating sensitivity to the UV cutoff
procedure and they will generate one-forms that are both closed and exact, {\it i.e.}
elements of the trivial cohomology of the Weyl group. Genuine anomalies
are clearly not in this class, since if they could be changed or removed by
adjustment of the UV counterterms then they would in no way be `anomalous.'
However, at the physical dimension it may happen that some combination(s) of the
local curvature invariants become Weyl invariant. This means that the Weyl derivation
by $\Delta_{\sigma}$ of the $d$ dimensional counterterm $S_d[g]$ must contain at
least one factor of $d-2k$, so that the finite limit,
\begin{equation}
\Gamma[\bar g;\sigma] =
\lim_{d\rightarrow 2k} {S_d[\bar ge^{2\sigma}] -  S_d[\bar g]\over d-2k}
=  \lim_{d\rightarrow 2k} {\Delta_{\sigma}\circ S_d[\bar g]\over d-2k}\,,
\label{dimlim}
\end{equation}
exists in the physical dimension $d=2k$. Whenever this limit exists the
resulting functional automatically satisfies the WZ consistency condition
(\ref{WZ}), since it is constructed explicitly as an exact one-form of the
Weyl group in $d$ dimensions, and the nilpotency property (\ref{nilp}) is
purely algebraic, independent of $d$, commuting with the limit in
(\ref{dimlim}). However, as we shall see, {\it after} the indicated limit
has been taken, the effective action $\Gamma[\bar g;\sigma]$ can no longer
be written as the Weyl variation of a local action in the physical dimension,
and hence each non-zero counterterm $S_d$ for which the limit (\ref{dimlim})
exists will generate an element of the WZ consistent non-trivial cohomology
of the Weyl group in precisely $d=2k$ dimensions. Conversely, since the anomaly
$\cal A$, is generated by taking the infinitesimal Weyl variation of the
effective action $\Gamma[\bar g;\sigma]$ via (\ref{AWZ}), and it
is composed of local dimension $2k$ scalar invariants with finite coefficients,
by taking the full set of these invariants near the physical dimension and finding
all conformal invariant combinations, all non-trivial cocycles of the Weyl
group must be generated in dimensional regularization by counterterms in the effective
action whose UV pole singularity at the physical dimension is cancelled by a
$d-2k$ factor in the numerator. Only in this way can a local anomaly in the trace
of the renormalized energy-momentum tensor with a finite UV cutoff independent
coefficient be generated. Hence dimensional regularization naturally classifies
the dimension $2k$ scalar invariants into those corresponding to non-trivial
and trivial first cocycles of the Weyl group, depending on whether their Weyl
variations do or do not vanish linearly as $d\rightarrow 2k$. There do not
appear to be invariants which vanish faster than linearly as $d\rightarrow 2k$,
which would correspond to higher non-trivial cohomological structures in the
space of metrics.

These general observations reduce the problem of finding all the
non-trivial elements of the cohomology of the Weyl group in $2k$
even dimensions to finding all the local Weyl invariants in that
dimension. This latter problem of classical differential geometry
has been solved by the construction of Fefferman and Graham which
we review in Section 6. Since the counterterms in dimensional
regularization are integrals over local invariants, we must allow
also for the possibility of non-trivial cocycles arising from
local densities that are Weyl invariant only up to total
derivatives. Such counterterms correspond to topological
invariants, of which there is exactly one, the Gauss-Bonnet-Euler
invariant $E_{2k}$ in any even dimension \cite{diffgeo}. This one
topological invariant gives rise to the type A anomalies while the
FG local Weyl invariants give rise to the type B anomalies
\cite{DesSch}. Both have Weyl variations that vanish linearly as
$d\rightarrow 2k$ and hence the continuation of each of these two
kinds of invariants away from $d=2k$ determines a finite WZ action
$\Gamma[\bar g;\sigma]$ independent of the UV regulator pole for
which the limit in (\ref{dimlim}) exists.

In this way the explicit construction of the non-trivial cohomology of the
Weyl group can be carried out, determining the general form of the conformal
anomaly and the non-local WZ consistent effective action corresponding to it
in any even dimension. In the following two sections we show that for $d=2$ and
$d=4$ dimensions this construction generates precisely the Polyakov action
and the four dimensional anomalous action analogous to it discussed in
previous work \cite{Poly,Rieg,AMa}. Generalizations to higher even dimensions are also
straightforward. Odd dimensional spacetimes have no conformal anomalies and must be
treated differently, {\it e.g.} as boundaries of spacetimes of one higher
dimension.

\section{WZ Action and Energy-Momentum in Two Dimensions}

Let us illustrate our general approach first in $d=2$ dimensions.
The unique dimension-two local scalar function of the metric is
the Ricci scalar, and hence its spacetime integral is the
only possible counterterm near $d=2$ dimensions. Thus we consider
\begin{equation}
\Gamma[\bar g;\sigma] = \lim_{d\rightarrow 2} {\int\,d^d\,x\,
\sqrt{-g}\, R - \int\,d^d\,x\,\sqrt{-\bar g}\,\overline R\over d-2}\,,
\label{Gamtwo}
\end{equation}
where $R= R[g]$ is the Ricci scalar in $d$ dimensions evaluated
on the $d$-dimensional metric $g_{ab} = e^{2\sigma} \bar g_{ab}$,
and $\overline R = R[\bar g]$, evaluated on the $d$-dimensional
fiducial metric $\bar g_{ab}$. Now in $d$ dimensions,
\begin{equation}
\sqrt{-g}\,R = \sqrt{-\bar g}\,
e^{(d-2)\sigma}\left[ \overline R - 2(d-1)\sqb \sigma - (d-1)(d-2)
\sigma^a\sigma_a\right]\,,
\label{Rd}
\end{equation}
where all covariant derivatives and contractions are performed
with the metric $\bar g_{ab}$ and we have introduced the shorthand notations,
$\overline\nabla_a\sigma \equiv \sigma_a$, $\sqb \sigma \equiv
\overline\nabla_a\overline\nabla^a \equiv \sigma^a_{\ ;a}$.
Expanding (\ref{Rd}) to first order in $d-2$, subtracting $\sqrt{-\bar
g}\,\overline R$  and taking the limit indicated in (\ref{Gamtwo}), we obtain
\begin{equation}
\Gamma[\bar g;\sigma] = \int\,d^2\,x\,\sqrt{-\bar g} \left[-\sigma\,\sqb\,
\sigma + \sigma\,\overline R\right]\,,
\label{poly}
\end{equation}
after an integration by parts and ignoring total derivatives which give
possible surface contributions. Up to a multiplicative normalization this is
exactly the Polyakov action found by functionally integrating with respect to
$\sigma$ the form of the trace anomaly for the metric $g_{ab} = \bar g_{ab}
e^{2\sigma}$ directly in $d=2$ dimensions \cite{Poly}, {\it i.e.}
\begin{equation}
{\delta \Gamma[\bar g;\sigma] \over \delta \sigma} =
\sqrt{-\bar g}\, (\overline R - 2\sqb \sigma) = \sqrt{-g} R\,.
\label{vartwo}
\end{equation}
When the conformal property of the self-adjoint hermitian differential operator, $\sq$
in two dimensions,
\begin{equation}
\sqrt{-g}\,\sq = \sqrt{\bar g} \,\sqb\,,\qquad (d=2)
\label{lap}
\end{equation}
is used, together with (\ref{Rd}) for $d=2$, it is easily checked that the
effective action $\Gamma [\bar g;\sigma]$ in (\ref{poly}) obeys the WZ consistency
condition (\ref{WZ}). Indeed, this is automatic from its construction (\ref{Gamtwo})
as an exact one-form in $d \neq 2$ dimensions, followed by a limiting process which
commutes with the algebraic nilpotency property of $\Delta_{\sigma}$.

However, since (\ref{poly}) is a simple polynomial in $\sigma$ it
cannot be written as a single-valued local functional of the full metric
$\bar ge^{2\sigma}$. This is clear from the fact that the complex transformation,
\begin{equation}
\sigma \rightarrow \sigma + i\pi q
\label{shift}
\end{equation}
for any integer $q$ leaves the full metric $g = \bar ge^{2\sigma}$ invariant
but under this transformation,
\begin{equation}
\Gamma[\bar g;\sigma] \rightarrow \Gamma[\bar g;\sigma] + iq\pi
\int\,d^2\,x\,\sqrt{-\bar g}\ \overline R = \Gamma[\bar g;\sigma]
+ 4{\pi}^2 iq \,\bar\chi_{_E} \label{multi}
\end{equation}
where $\bar\chi_{_E}$ is the Euler number of the metric $\bar g$.
Hence the action $\Gamma[\bar g;\sigma]$ is not a single-valued
local functional of $g= \bar ge^{2\sigma}$, but rather an explicit representation
of the non-trivial cohomology of the Weyl group in $d=2$ physical dimensions.
Evidently, the expansion of the exponential conformal $\sigma$ dependence
of (\ref{Rd}) required by the limit (\ref{Gamtwo}) is responsible for $\Gamma$
being a simple polynomial in $\sigma$ without the periodicity under (\ref{shift}),
and this is associated in turn with the fact that the integral of the Ricci
scalar is invariant under the Weyl group only in exactly two dimensions.
Since the integral of $R$ is proportional to the Euler number in two dimensions,
the existence of a single element of the non-trivial cohomology and a single
anomaly coefficient is a consequence of the existence of one and only one (type
A) topological invariant. There are no (type B) local Weyl invariants in two
dimensions.

Notice also that if instead of $q$ being an integer we take $i\pi q$ to be a
real constant $\sigma_0$, then the transformation (\ref{shift}) is
a global rescaling of the metric,
\begin{equation}
g_{ab} \rightarrow g_{ab}e^{2\sigma_0}\,.
\label{global}
\end{equation}
Then (\ref{multi}) informs us that the multi-valuedness of the action
$\Gamma[\bar g;\sigma]$ necessarily implies that it transforms non-trivially under such
a global rescaling. Hence, non-trivial behavior under global rescaling is a necessary
feature of a representative of the non-trivial cohomology of the Weyl group.

Associated with the one-form $\Gamma$, which is local in $\sigma$,
obeying WZ consistency there is a non-local zero form, {\it i.e.} a
non-local quantum effective action of the full metric $g= \bar ge^{2\sigma}$.
Since the action (\ref{poly}) is quadratic in $\sigma$ this non-local
action is easily constructed by adding to $\Gamma$ a $\sigma$ independent
piece which `completes the square' and leaves the variation (\ref{vartwo})
and the WZ condition unaffected. To find this $\sigma$ independent term
explicitly, we need only solve the linear differential equation (\ref{vartwo})
for $\sigma$ by introducing the Green's function inverse of the second order
Weyl covariant differential operator (\ref{lap}), namely
\begin{equation}
-\sqrt{-g}\,\sq\, D_2(x,x') = \delta^2 (x,x')\,,
\label{inv}
\end{equation}
where $\delta^2 (x,x')$ is a density of weight two, {\it i.e.}
its integral $\int d^2x'\,\delta^2 (x,x') = 1$. If the metric
has a Lorentzian signature then the boundary conditions needed
to specify the particular Green's function solution of (\ref{inv})
will depend on the application. If $\sq$ has normalizable zero
modes, such as on the Euclidean two sphere $S^2$, then $D_2(x,x')$
will have to be defined by the inverse of $\sq$ on the orthogonal
complement to its kernel, the $\delta^2$ function on the right side
of (\ref{inv}) being appropriately modified.
Using the Green's function, $D_2$ we obtain from (\ref{vartwo}),
\begin{equation}
\sigma (x) = {1\over 2}\int\,d^2\,x'\, D_2(x,x')\left(\sqrt{-g}\,R -
\sqrt{-\bar g}\,\overline R \right)_{x'}\,\qquad (d=2)\,.
\label{soltwo}
\end{equation}
This is a formal solution of the Poincare uniformization problem
in two dimensions, since the metric $\bar g_{ab}$ may be chosen as
that of constant scalar curvature $\overline R$, and then
(\ref{soltwo}) specifies the conformal factor needed to bring an
arbitrary metric $g_{ab}$ (with the same fixed Euler number as
$\bar g_{ab}$) to the metric with uniform scalar curvature
\cite{Yamabe}.

Substituting the solution for $\sigma$ (\ref{soltwo}) into
(\ref{poly}) and using the fact that $D_2(x,x') = \overline
D_2(x,x')$ is Weyl invariant (owing to the Weyl invariance of
$\sqrt{-g}\,\sq$) gives
\begin{equation}
\Gamma[\bar g;\sigma] = {1\over 4} \int\,d^2\,x\,\sqrt{-g}
\int\,d^2\,x'\,\sqrt{-g'}\, R(x)\,D_2(x,x')\,R(x')
- {1\over 4} \int\,d^2\,x\,\sqrt{-\bar g}
\int\,d^2\,x'\,\sqrt{-\bar g'}\, \overline R(x)\,\overline D_2(x,x')\,\overline
R(x')\,. \label{Gamint}
\end{equation}
This difference form shows again that $\Gamma[\bar g;\sigma]$ is the Weyl variation
$\Delta_{\sigma}$ of a scalar action functional (although now non-local) in the physical
dimension $d=2$, which accounts for it automatically satisfying the WZ condition.
Clearly, the multi-valuedness of this functional (\ref{multi}) is related to
its non-locality by the fact that it depends on $\sigma$ directly
(rather than $e^{2\sigma}$) and eliminating $\sigma$ by (\ref{soltwo})
introduces the non-local Green's function $D_2(x,x')$. This example
illustrates how the non-trivial cohomology of the Weyl group is associated
with non-local action functionals in the physical dimension, despite the fact
that we began with a local functional in (\ref{Gamtwo}) defined in $d$ dimensions.

The last term in (\ref{Gamint}) is precisely the $\sigma$ independent
term required to bring $\Gamma$ into the fully covariant but non-local form.
Defining $\Gamma_{WZ}$ by multiplying $\Gamma$ by the correct normalization
factor of $-Q^2/4\pi \equiv c/24\pi$, corresponding to
the standard two-dimensional trace anomaly coefficient $c_m = N_s + N_f$
for $N_s (N_f)$ free massless scalar (fermion) matter fields,
we obtain
\begin{equation}
\Gamma_{WZ}[\bar g;\sigma] = \Delta_{\sigma}\circ S_{anom}[\bar g]
= S_{anom}[g] - S_{anom}[\bar g]\,, \end{equation}
with
\begin{equation}
S_{anom}[g] = -{Q^2\over 16\pi} \int\,d^2\,x\,\sqrt{-g}
\int\,d^2\,x'\,\sqrt{-g'}\, R(x)\,D_2(x,x')\,R(x')\,,
\label{acttwo}
\end{equation}
which is the fully covariant non-local form of the Polyakov
action in two dimensions \cite{Poly}. The total central charge is $c=c_m -26 +1$
when the effects of ghosts and the $\sigma$ field itself are included
in the trace anomaly coefficient and $Q^2 = (25-c_m)/6$.

In this way the non-local WZ consistent effective action (\ref{acttwo})
corresponding to the trace anomaly in two dimensional spacetime can be
constructed from the counterterm in dimensional regularization
near $d=2$ dimensions. Since $\int\,d^2\,x\,\sqrt{-g}\, R = 4\pi \chi_{_E}$
is the unique integral Weyl invariant in two dimensions, there are no local UV
counterterms of dimension two which can be added to the effective action and
(\ref{acttwo}) cannot be removed or altered by any UV counterterm.
On the contrary, the anomaly calculation shows that its coefficient is
determined by the number of massless excitations in the far infrared,
which is a genuine feature of the low energy theory of $2D$ gravity,
independent of any UV regulator. This physical meaning of the non-trivial
cohomology of the Weyl group in two dimensions through the non-local
IR effective action (\ref{acttwo}) and its properties under global rescalings
has been verified by simplicial lattice simulations \cite{DTs,simp}.

Corresponding to the covariant non-local effective action (\ref{acttwo})
there is a conserved energy-momentum tensor. The most rapid route to deriving
this tensor is to vary the local form of the WZ action with respect to the
background metric $\bar g$. In this way we obtain
\begin{equation}
T_{ab}^{WZ}[\bar g;\sigma] \equiv -{2\over \sqrt{-\bar g}}
{\delta \Gamma_{WZ}[\bar g;\sigma] \over \delta \bar g^{ab}} =
-{Q^2\over 4\pi}\left[2\sigma_{;ab}
-2 \bar g_{ab} \sqb\sigma - 2 \sigma_a\sigma_b
+ \bar g_{ab}\, \sigma_c\sigma^c\right]\,,
\label{enerpoly}
\end{equation}
where the shorthand notation $\sigma_{;ab} \equiv \overline\nabla_b
\overline\nabla_a \sigma = \overline\nabla_a\overline\nabla_b\sigma$
has been employed.

It is easily checked that the same result is obtained
from varying the $d$-dimensional local counterterm in (\ref{Gamtwo}).
This variation is proportional to the difference of Einstein tensors:
$G_{ab}-\overline G_{ab}$ which vanishes identically at $d=2$.
However, in $d \neq 2$ dimensions,
\begin{equation}
G_{ab} = \overline G_{ab} + (d-2)\left[-\sigma_{;ab}
+ \bar g_{ab}\, \sqb\sigma + \sigma_a\sigma_b
+ \left({d-3\over 2}\right) \bar g_{ab}\, \sigma_c\sigma^c
\right]\,.
\end{equation}
Inserting this into the variation of (\ref{Gamtwo}), taking
account of the factor of $-2$ in the definition of the stress tensor in
(\ref{enerpoly}) and taking the limit to the physical dimension $d=2$ gives
(\ref{enerpoly}) again. This calculation shows that the conserved stress
tensor of the Polyakov action is related in fact to the conformal variation of
the Einstein tensor away from $d=2$ dimensions, a result that was
anticipated in earlier considerations of the
renormalized stress tensor in dimensional regularization \cite{Bunch}.

Finally we note that since the solution for $\sigma$, (\ref{soltwo}) can
be written as the difference of two terms evaluated on the metrics
$g_{ab}$ and $\bar g_{ab}$ respectively, we can introduce this solution
into (\ref{enerpoly}). Using the identities of Appendix A relating the
covariant derivatives with respect to the two metrics, we find that
the mixed terms cancel and the stress tensor may be written as the
difference of two tensors, $T_{ab}^{WZ} = T_{ab}^{(anom)}[g] - T_{ab}^{(anom)}
[\bar g]$, each of which is a non-local function of a single metric. Indeed,
\begin{equation}
T_{ab}^{(anom)}[g] \equiv -{2\over \sqrt{- g}} {\delta S_{anom}[g]
\over \delta g^{ab}} = {Q^2\over 4\pi}\left[-\nabla_a\nabla_b\varphi
+ g_{ab}\, \sq\varphi - {1\over 2}(\nabla_a\varphi)(\nabla_b\varphi)
+ {1\over 4} g_{ab}\, (\nabla_c\varphi)(\nabla^c\varphi)\right]\,,
\end{equation}
where
\begin{equation}
\varphi (x) \equiv \int\,d^2\,x'\, D_2(x,x')\sqrt{-g'}\,R'\qquad (d=2)\,.
\label{phitwo}
\end{equation}
It is easily checked that the non-local tensor $T_{ab}^{anom}[g]$ is conserved
by virtue of the vanishing of the Einstein tensor in $d=2$ dimensions, and that
it has the trace,
\begin{equation}
g^{ab}\,T_{ab}^{(anom)} = {Q^2\over 4\pi} \sq \varphi
=- {Q^2\over 4\pi}\, R = {c\over 24\pi}\, R\,,
\end{equation}
which is the local trace anomaly in two dimensions.

\section{The WZ Action in Four Dimensions}

In $d=4$ spacetime dimensions there are four local scalar functions
of the metric with dimension four, {\it viz.} $R^2$, $R_{ab}R^{ab}$,
$R_{abcd}R^{abcd}$, and $\sq R$. The last of these is a total
derivative in any number of dimensions, so it gives no volume contribution
when integrated. Thus there are just three possible counterterms for the
dimensionally continued effective action for gravity near four dimensions.
However, the existence of the total derivative $\sq R$ indicates a new
feature absent in $d=2$, since $\sq R$ can appear in the trace anomaly but
disappears from the volume effective action. This is associated with the existence
of a trivial cocycle in four dimensions and leads to one constraint
between the four possible terms in the trace anomaly \cite{duffa,Bon}.

Let us define the two linear combinations in $d$ near equal to four dimensions,
\begin{mathletters}
\begin{eqnarray}
E_d &\equiv& R_{abcd}R^{abcd} -4R_{ab}R^{ab} + R^2 + {(d-4)\over
18}R^2\,;\quad {\rm and} \label{EFdef}\\
F_d &\equiv&[C_{abcd}C^{abcd}]_d = R_{abcd}R^{abcd}
-{4\over d-2}R_{ab}R^{ab}  + {2\over (d-1)(d-2)}R^2\nonumber\\
&=& F_4 + {(d-4) \over 18}\, R^2 + (d-4)\left(R_{ab}R^{ab} - {1\over 3}R^2\right)
+ {\cal O} (d-4)^2\,,
\end{eqnarray}
\end{mathletters}
\noindent which together with $R^2$ form a basis for the three remaining independent 
scalar
invariants in the effective action. At $d=4$, $E_4$ is the integrand of the
Gauss-Bonnet-Euler topological invariant, analogous to $R$ at $d=2$, while
$F_4$ is the Weyl tensor squared, a local Weyl invariant. Each of these
is Weyl invariant in $d=4$ when integrated over all space. Thus, each
of these two terms will generate a non-trivial cocycle of the Weyl group in $d=4$.
The addition of the $R^2$ term with the particular coefficient $(d-4)/18$ in
(\ref{EFdef}) adds a particular admixture of the trivial cocycle in defining the
$E_d$ invariant away from $d=4$, chosen with a view ahead to simplify the
$d\rightarrow 4$ limit. We do not add any such term to $F_d$ since it already
transforms as a local density of weight $4$ under the Weyl group, and
\begin{equation}
\sqrt{-g} F_d =
\sqrt{-\bar g}\,e^{(d-4)\sigma} \overline F_d =  \sqrt{-\bar g}\,\overline
F_d + (d-4)\sigma \sqrt{-\bar g}\,\overline F_d + {\cal O}(d-4)^2\,.
\label{varF} \end{equation}
becomes a local Weyl invariant in exactly $d=4$ dimensions.

Our general algorithm for the construction of the WZ consistent effective action
over the two non-trivial cocycles now requires that we evaluate
\begin{equation}
\Gamma_{WZ}[\bar g;\sigma] = b \lim_{d\rightarrow 4} {\int\,d^d\,\,x\,
\sqrt{-g}\, F_d - \int\,d^d\,\,x\,\sqrt{-\bar g}\,\overline F_d\over d-4}
+ b'\lim_{d\rightarrow 4} {\int\,d^d\,\,x\,
\sqrt{- g} E_d - \int\,d^d\,\,x\,\sqrt{-\bar g}\,\overline E_d\over d-4}\,,
\label{Gamfour}
\end{equation}
with arbitrary coefficients $b$ and $b'$. Expanding the simple transformation
law (\ref{varF}) to linear order in $d-4$ immediately gives the form of the
first non-trivial cocycle, namely
\begin{equation}
b \lim_{d\rightarrow 4} \left\{{\int\,d^d\,x\,\sqrt{-g}\,F_d -
\int\,d^d\,x\,\sqrt{-\bar g}\,\overline F_d\over d-4}\right\} =
b  \int\, d^4x\,\sqrt{-\bar g}\,\overline F_4\,\sigma \,,
\label{Fcocycle}
\end{equation}
which is linear in $\sigma$.

The algebra required to compute the second cocycle in (\ref{Gamfour}) is somewhat
more tedious. The necessary relations are cataloged in
Appendix A. To simplify the task one may note first that
\begin{equation}
E_d = F_d -\left[ 2 + (d-4)\right]\left(R_{ab}R^{ab} - {1\over 3}R^2\right)
+ {\cal O}(d-4)^2\,,
\label{EFrel}
\end{equation}
so that the only non-trivial quantity whose $\sigma$ dependence we need near
four dimensions is $R_{ab}R^{ab} - {1\over 3}R^2$. Using
\begin{equation}
R_{ab} = \overline R_{ab} - (d-2)(\sigma_{;ab} -\sigma_a\sigma_b + \bar
g_{ab}\, \sigma_c\sigma^c) - \bar g_{ab}\, \sigma_{;c}^c\,,
\end{equation}
and (\ref{Rd}) we show in Appendix A that
\begin{eqnarray}
&& \int\,d^d\,x \sqrt{-g}\,\left(R_{ab}R^{ab} - {1\over 3}R^2\right) =
\int\,d^d\,x \sqrt{-\bar g}\,\left(\overline R_{ab}\overline R^{ab} - {1\over 3}\overline
R^2\right)
\nonumber\\
&&\qquad + (d-4) \int\,d^d\,x \sqrt{-\bar g}\, \sigma  \left(\overline
R_{ab}\overline R^{ab} - {1\over 3}\overline R^2 + {1\over 3}\sqb \overline
R\right) -(d-4)\int\,d^d\,x \sqrt{-\bar g}\,  \sigma\,\bar\Delta_4\, \sigma +
{\cal O}(d-4)^2\,,
\label{Rexp}
\end{eqnarray}
up to surface terms which we systematically neglect. In this expression
\begin{equation}
\Delta_4 \equiv \sq^2 + 2 R^{ab}\nabla_a\nabla_b +{1\over 3} (\nabla^a R)
\nabla_a - {2\over 3} R \sq
\label{Deldef}
\end{equation}
is the fourth order scalar operator satisfying the Weyl invariance property
in four dimensions,
\begin{equation}
\sqrt{-g} \Delta_4 = \sqrt{-\bar g} \bar \Delta_4\qquad (d=4)\,,
\label{invfour}
\end{equation}
analogous to (\ref{lap}) in two dimensions.
The cancellation of all terms cubic and quartic in $\sigma$ which {\it a priori}
could appear in the last line of (\ref{Rexp}) is noteworthy.
Using (\ref{Rexp}) we have
\begin{eqnarray}
&&\int\,d^d\,x\, \sqrt{-g}\,E_d  = \int\,d^d\,x\, \sqrt{-g}\,F_d
-[2 + (d-4)] \int\,d^d\,x\, \sqrt{-g}\,\left(R_{ab}R^{ab} - {1\over 3} R^2\right)
+ {\cal O}(d-4)^2 \nonumber\\
&& \qquad =\int\,d^d\,x\, \sqrt{-\bar g}\,\overline F_d + (d-4)
\int\,d^d\,x\,\sqrt{-\bar g}\,
\overline F_d\,\sigma - [2 + (d-4)] \int\,d^d\,x\,\sqrt{-\bar g}\,
\left(\overline R_{ab}\overline R^{ab} - {1\over 3} \overline R^2 \right)
\nonumber\\
&&\qquad\qquad - 2 (d-4) \int\,d^d\,x\,\sqrt{-\bar g}\,  \left(\overline R_{ab}\overline
R^{ab}
- {1\over 3} \overline R^2 + {1\over 3} \sqb \overline R\right)\, \sigma
+ 2 (d-4) \int\,d^d\,x\,\sqrt{-\bar g}\,\sigma\,\bar\Delta_4\, \sigma
+ {\cal O}(d-4)^2 \nonumber\\
&&\qquad =\int\,d^d\,x\, \sqrt{-\bar g}\,\overline E_d
+ (d-4) \int\,d^d\,x\, \sqrt{-\bar g}\,\left\{\left(\overline E_d - {2\over 3}\sqb
\overline R\right)\,\sigma + 2 \sigma\,\bar\Delta_4\, \sigma\right\}
+ {\cal O}(d-4)^2\,.
\end{eqnarray}
Therefore, neglecting possible surface terms we find that the second term in
(\ref{Gamfour}) becomes
\begin{equation}
b' \lim_{d\rightarrow 4} \left\{{\int\,d^d\,x\, \sqrt{-g}\,E_d -
\int\, d^d\,x\,\sqrt{-\bar g}\,\overline E_d\over d-4}\right\} =
b' \int\, d^4x\, \sqrt{-\bar g}\,\left\{\left(\overline E_4 - {2\over 3}
\sqb \overline R\right)\sigma + 2\,\sigma\bar\Delta_4\sigma\right\}\,,
\end{equation}
and the general element of the non-trivial cohomology of the
Weyl group in four dimensions is given by
\begin{equation}
\Gamma_{WZ}[\bar g;\sigma] = b  \int\,d^4x\,\sqrt{-\bar g}\, \overline F_4\,\sigma
+ b' \int\,d^4x\,\sqrt{-\bar g}\,\left\{\left(\overline E_4 - {2\over 3}
\sqb \overline R\right)\sigma + 2\,\sigma\bar\Delta_4\sigma\right\}\,.
\label{genfourWZ}
\end{equation}
Notice that this construction of $\Gamma_{WZ}$ contains only terms up to
quadratic order in $\sigma$, which was arranged by the addition of the local
$R^2$ term with the particular coefficient $(d-4)/18$ in (\ref{EFdef}).

In exactly $d=4$ dimensions by using the invariance property (\ref{invfour})
of the self-adjoint hermitian differential operator $\Delta_4$,
it is easily checked that the Weyl one-form $\Gamma_{WZ}[\bar g;\sigma]$
satisfies the WZ consistency condition, as was first shown in ref. \cite{AMMnp}.
In the present treatment this follows immediately from the construction
of (\ref{Gamfour}) as the limit of a $d$ dimensional exact form.
The $\sigma$ variation of $\Gamma_{WZ}[\bar g;\sigma]$,
\begin{eqnarray}
{\delta \Gamma_{WZ}[\bar g;\sigma]\over \delta \sigma} &=&
b \sqrt{-\bar g}\,\overline F_4 + b' \sqrt{-\bar g}\,
\left\{\left(\overline E_4 - {2\over 3}
\sqb \overline R\right)\sigma + 4\,\sigma\,\bar\Delta_4\,\sigma\right\}
\nonumber\\
&=& b \sqrt{-g}\,F_4 + b' \sqrt{-g}\,\left(E_4 - {2\over 3}
\sq R\right)\,,
\label{Gamvar}
\end{eqnarray}
is the non-trivial conformal trace anomaly of massless quantum
matter fields in the full metric, $g_{ab} = \bar g_{ab} \exp(2\sigma)$,
as is also easily checked using the conformal variation formulae
derived in Appendix A, since
\begin{equation}
\sqrt{-g}\,\left(E_4 - {2\over 3}\sq R\right) = \sqrt{-\bar g}\,
\left(\overline E_4 - {2\over 3}\sqb\overline R\right) + 4\,\sqrt{-\bar g}\,
\bar\Delta_4\,\sigma\,.
\label{Esig}
\end{equation}
Using this relation for $g_{ab}$ and $\bar g_{ab}$ interchanged and
$\sigma \rightarrow - \sigma$ is the most immediate way of proving the
invariance property (\ref{invfour}). The independent construction of
$\Gamma_{WZ}[\bar g;\sigma]$ by the dimensional continuation limiting process
(\ref{Gamfour}) explains why the action obtained by integration of the trace
anomaly by undoing the variation (\ref{Gamvar}) gives a WZ consistent
effective action for the local conformal factor of the metric in four
dimensions \cite{AMMnp}.

As in two dimensions, the fact that $\Gamma_{WZ}[\bar g;\sigma]$ is a
simple second order polynomial in $\sigma$ means that it cannot be
written as a single-valued local functional of the full metric
$g_{ab}$, although its $\sigma$ variation can be through (\ref{Gamvar}).
One can again consider the complex transformation (\ref{shift}) and
observe that
\begin{equation}
\Gamma_{WZ}[\bar g;\sigma] \rightarrow \Gamma_{WZ}[\bar g;\sigma]
+ i\pi q b \int\, d^4x\,\sqrt{-\bar g}\,\overline F_4 + 32\pi^3 iq
b' \bar \chi_{_E}\,, \label{scafour}
\end{equation}
so that neither term is single valued. Replacing $i\pi q$ by
a constant real shift $\sigma_0$ shows that this multi-valuedness
of the action $\Gamma_{WZ}[\bar g;\sigma]$ associated with its
non-trivial cohomology necessarily implies its sensitivity to global Weyl
rescalings of the metric. Neither term of the non-trivial cocycle is invariant
under global dilations, contrary to what is claimed in ref. \cite{DesSch} for
the type A anomaly, and both terms in the WZ consistent effective action arise
from the same mechanism through a cancellation of the pole in dimensional regularization.
It is also noteworthy that the admixture of the $\sq R$ term in the second
cocycle drops out of the global dilation (\ref{scafour}), since it is a total
derivative. The sensitivity to finite volume scaling comes only from the
Weyl invariants $F_4$ and $E_4$ in four dimensions.

Further, we can exhibit the non-local but fully covariant form
of the WZ effective action by introducing the Green's function
of the fourth order Weyl covariant differential operator (\ref{Deldef}).
Defining this Green's function, $D_4$ by
\begin{equation}
\sqrt{-g}\Delta_4 D_4(x,x') = \delta^4(x,x)\,,
\end{equation}
with the same qualifying remarks as follow (\ref{inv}), allows
us formally to invert the relation (\ref{Esig}) for $\sigma$ to obtain
\begin{equation}
\sigma (x) = {1\over 4} \int\,d^4x'\,D_4(x,x')\left[
\sqrt{-g}\,\left(E_4 - {2\over 3} \sq R\right) -
\sqrt{-\bar g}\, \left(\overline E_4 - {2\over 3} \sqb \overline R\right)
\right]_{x'}\,.
\label{solfour}
\end{equation}
This is a formal solution to the four dimensional uniformization
problem of bringing an arbitrary metric to one with constant $E_4
- {2\over 3} \sq R$ by a local Weyl conformal transformation. We
remark that although the Poincare-Yamabe conjecture has been
proven in two dimensions \cite{Yamabe}, this four dimensional
uniformization conjecture (namely, that the $\sigma$ of our formal
inversion of $\Delta_4$ exists and is unique) has not been proven.
However, the conformal property of $\Delta_4$ suggests that it
should be possible to generalize the two dimensional case in this
way.

Substituting (\ref{solfour}) into (\ref{genfourWZ})
and using $D_4 (x,x') =\overline D_4(x,x')$ shows
that $\Gamma_{WZ}[\bar g;\sigma]$ can be written explicitly as
a difference of non-local actions,
\begin{equation}
\Gamma_{WZ}[\bar g;\sigma] = \Delta_{\sigma}\circ S_{anom}[\bar g]
= S_{anom}[g] - S_{anom}[\bar g]\,,
\label{anomdiff}
\end{equation}
with
\begin{eqnarray}
S_{anom}[g] &=& {b \over 4} \int d^4x\,\sqrt{-g}\, \int d^4x'\,\sqrt{-g'}\,
F_4\, D_4(x,x') \left(E_4 - {2\over 3} \sq R\right)' \nonumber\\
+&&
{b'\over 8} \int d^4x\,\sqrt{-g}\, \int d^4x'\,\sqrt{-g'}\,
\left(E_4 - {2\over 3} \sq R\right)\, D_4(x,x')
\left(E_4- {2\over 3} \sq R\right)'\,.
\label{anomact}
\end{eqnarray}
If we introduce the auxiliary field $\varphi$ defined in $d=4$ by
\begin{equation}
\varphi (x) \equiv {1\over 2} \int\,d^4x'\,D_4(x,x')
\sqrt{-g'}\,\left(E_4 - {2\over 3} \sq R\right)'\qquad (d=4)\,,
\label{phifour}
\end{equation}
then the non-local action may be written in the form,
\begin{equation}
S_{anom}[g] = {b \over 2} \int d^4x\,\sqrt{-g}\,
F_4\, \varphi - {b'\over 2} \int d^4x\,\sqrt{-g}\, \left[\varphi\, \Delta_4\,
\varphi - \left(E_4- {2\over 3} \sq R\right)\,\varphi \right]
\equiv S_{anom}^{(F)} + S_{anom}^{(E)}
\label{actphi}
\end{equation}
The result (\ref{anomact}) corrects a misprint in eq. (2.14) of ref.\cite{AMa}.

In this way the non-local WZ consistent covariant effective action
corresponding to the non-trivial cohomology of the Weyl group in four
dimensions may be constructed directly from Weyl invariant counterterms
in dimensional regularization. The trivial cocycle corresponds to
the Weyl non-invariant local action,
\begin{equation}
S_{local}^{(4)}[g] = -{(2b + 2b' + 3b'')\over 18}\int d^4x\,\sqrt{-g}\, R^2\,,
\label{localSf}
\end{equation}
to conform to Duff's parameterization of the three independent trace anomaly
coefficients \cite{duffa}, namely
\begin{equation}
T_a^{\ a}= {\cal A}_4 = {1\over \sqrt{-g}} {\delta \over \delta
\sigma}(S_{anom} + S_{local}^{(4)}) = b\left(F_4 + {2\over 3}\sq R\right) + b' E_4 +
b'' \sq R\,.
\label{anomD}
\end{equation}
Whereas the coefficients $b$ and $b'$ are determined by the number of massless
fields of each spin, the $b''$ coefficient is scheme dependent, as we expect
for the true UV counterterm $\int d^d\,x \sqrt{-g}\,R^2$ that has non-vanishing
local Weyl variation in the physical dimension $d=4$ and no cancellation of the
$(d-4)^{-1}$ pole multiplying it.

The local $R^2$ invariant which is a trivial cocycle of the Weyl group
differs from the non-trivial $F$ and $E$ cocycles in another way. As we have
seen, these are sensitive to global Weyl rescalings $\sigma \rightarrow \sigma
+ \sigma_0$ in the physical dimension, due to their multi-valuedness under the
complex periodic transformation of the metric $\sigma \rightarrow \sigma +
i\pi q$. This means that the non-trivial cocycles scale non-trivially
as the finite volume of the system is scaled. However, the trivial $R^2$
cocycle is single valued under the same global transformation. Indeed,
\begin{equation}
\sqrt{-g}\,R^2 = \sqrt{-\bar g}\,\left(\overline R -
6\sqb\sigma - 6\,\sigma_a\sigma^a \right)^2\qquad (d=4)\,,
\end{equation}
so its $\sigma$ dependence enters purely through derivatives, and its integral
is invariant under rigid global rescalings in the physical dimension. Hence
it is insensitive to finite volume rescalings, as one would expect for a
term relevant in the UV but irrelevant in the IR. The rigid dilation invariance
has associated with it a Noether current $J^a$, whose divergence $\nabla_a
J^a$ is proportional to the total derivative $\sq R$ in the anomaly. Since the
integral of the anomaly $\int d^4x \sqrt{-g}\, {\cal A}_4$ is nothing but the
global Weyl variation of the WZ effective action, in order to have vanishing
contribution to this global anomaly and therefore insensitivity to global
Weyl rescalings, trivial cocycles must yield total divergences in ${\cal A}_4$,
in distinction to both the non-trivial $F$ and $E$ cocycles. This clear
separation of behavior under global dilations shows that the trivial
and non-trivial cocycles of the Weyl group are associated with UV and
IR physics respectively.

We can proceed further to deduce a general property of the effective action
for gravity by classifying the behavior under the Weyl group. Since $\Gamma_{WZ}$
satisfying WZ consistency is unique up to an arbitrary
admixture of local trivial cocycles $\Delta_{\sigma}\circ S_{local}$, and
$\Gamma_{WZ}$ itself can be written as a finite Weyl shift on an anomalous
action, as in (\ref{anomdiff}), the only possible additions to $S_{anom}$ are
local terms or arbitrary (generally non-local) but Weyl invariant terms, $S_{inv}$
which drop out of difference (\ref{anomdiff}). That is, the full effective
action of any covariant theory must be of the form,
\begin{equation}
S_{eff}[g] = S_{local}[g] + S_{inv}[g] + S_{anom}[g]\,,
\label{totalS}
\end{equation}
where
\begin{equation}
\Delta_{\sigma}\circ S_{inv} = 0\,.
\end{equation}
In addition to (\ref{localSf}) the $S_{local}$ in this expression can contain
local terms of both higher and lower dimension than four, multiplied by coefficients
with negative or positive mass dimensions, respectively. The higher dimension terms are
strictly irrelevant in the IR, since they scale to zero with negative powers of
$e^{\sigma_0}$ and may be neglected for physics far below the Planck
scale, while the lower dimension local terms are nothing but the terms of the
usual Einstein-Hilbert classical action, {\it i.e.}
\begin{equation}
S_{local}[g] = {1\over 16\pi G}\int\,d^4\,x\,\sqrt{-g}\,(R-2\Lambda) +
S_{local}^{(4)} + \sum_{n=3}^{\infty} S_{local}^{(2n)}\,.
\label{locsum}
\end{equation}
The classical terms grow as positive powers of $e^{\sigma_0}$
under global dilations and are clearly IR relevant terms. The term
(\ref{localSf}) is the only allowed dimension four, Weyl
non-invariant local term. The local dimension four term involving
the Weyl tensor squared is Weyl invariant and among the many terms
that can appear in $S_{inv}$. Because both of these are neutral
under global dilations we expect them to be marginally irrelevant
in the IR (conversely, marginally relevant in the UV). All the
higher dimension local terms in the sum in (\ref{locsum}) for
$n\ge 3$ scale to zero as $\sigma_0 \rightarrow \infty$ and are
clearly strictly IR irrelevant. Any non-local, Weyl non-invariant
terms generate non-trivial cocycles of the Weyl group. If there
are only two non-trivial cocycles in $d=4$, then the most general
non-local, Weyl non-invariant action is given by (\ref{anomact}).
These scale linearly with $\sigma_0$ or equivalently,
logarithmically with length or volume rescalings. Although the
precise matter content of the massless fields which are integrated
out to obtain the effective gravitational action influence the
values of the $b$ and $b'$ coefficients, the form of $S_{anom}$ is
the completely general solution to the non-trivial cocycle action
in four dimensions and its response under global scale
transformations cannot be changed by local terms or Weyl invariant
terms.

Since the complete classification of terms in the effective action according to their
response under the Weyl group allows only local or completely Weyl invariant terms to be
added to $S_{anom}$, according to (\ref{totalS}), we can conclude definitively that there
is
{\it no} non-local term in $S_{eff}$ of the form,
\begin{equation}
\int\, d^4\,x\,\sqrt{-g}\,
C_{abcd}\,\log\left({\sq\over\mu^2}\right)\,C^{abcd}\,,
\label{wrongC}
\end{equation}
as has long been conjectured \cite{DDI,Desera,DesSch}. Indeed, such a term has
no simple transformation properties under the local Weyl group and its 
{\it local} Weyl variation does {\it not} generate either the $b$ or $b'$ term in the
local anomaly (\ref{anomD}). Although a plethora of complicated non-local terms
are generated in the full effective action of a quantum
field theory in curved spacetime, their Weyl non-invariant contributions
which are not either absorbable into redefinitions of the coupling
constants in the relevant parts of $S_{local}, n=0,1$ or strictly irrelevant,
($n\ge 3$) in the expansion (\ref{locsum}) are completely determined by the trace anomaly 
of the renormalized energy-momentum
tensor \cite{BarVil}. Therefore the effective action must always contain as
one piece, $S_{anom}[g]$, if its variation is to yield the correct local $F_4$
and $E_4$ trace anomalies in its conformal limit, and these non-trivial
Weyl cocycle terms cannot be removed or altered by the addition of local
terms. Since the term (\ref{wrongC}) is neither local nor
Weyl invariant, and it does not reduce to the terms in the non-trivial
cocycle that produce the correct local Weyl anomaly (by construction), it cannot appear 
in the effective action in a general background
metric (although it may mimic some effects of the correct $S_{anom}$ in the weak
field expansion around flat space). Although this conclusion and the decomposition
of the general form of the full low energy effective action of gravity in
four dimensions (\ref{totalS}) has been reached after lengthy calculations from the form
of the heat kernel expansion of the quantum effective action \cite{BarVil}, in fact
(\ref{totalS}) follows only from general covariance and the classification of terms in
the effective action according to their behavior under the Weyl group (\ref{weyl}).

As in $d=2$ the appearance of a locally Weyl covariant scalar differential operator
$\Delta_4$ is a necessary feature of the non-trivial cohomology. It cannot
be removed by the addition of local terms. The propagator of a conformal
differential operator $D_2$ or $D_4$ is a logarithm in coordinate space in any
number of dimensions. Its IR properties in position space do not become any worse in
higher dimensions than in $d=2$. In momentum space there is nothing unphysical about
either $\Gamma_{WZ}$ or $S_{anom}$, despite the appearance of the fourth order
differential operator. Although the apparent $p^{-4}$ pole produced some premature
concern \cite{Desera}, more careful attention to all the powers of momentum in the
numerator of flat space amplitudes shows that this concern is baseless and that
$S_{anom}$ is fully consistent with conformal Ward identities in flat space
\cite{Descor}. This certainly has to be the case without detailed calculation, since
the WZ consistent effective action is nothing but the generating functional of precisely
these conformal Ward identities. Any new non-local terms of the kind suggested
recently in \cite{Descor} are both not needed and inconsistent with the general
form (\ref{totalS}).

To conclude this section we remark also that canonical quantization of the quadratic
WZ action  $\Gamma_{WZ}$ on the cylindrical background Einstein space, $R \times S^3$
shows that there are {\it no} propagating ghost states which survive imposition of the
constraints of diffeomorphism invariance. Instead the physical Hilbert space of the
pure WZ theory ({\it i.e.} with the local terms $S_{local}$ set to zero)
consists of a particular global mode of the $S^3$ with a discrete
spectrum labelled by a single integer \cite{states}. This is exactly what one
would expect of a theory with a single IR degree of freedom, but no local UV degrees
of freedom (ghost or otherwise) such as occur in local higher derivative gravity
theories with $C_{abcd}C^{abcd}$ or $R^2$ actions. Those local higher
derivative actions certainly give rise corrections to Einstein gravity at
the Planck scale, where the entire low energy effective action approach
breaks down. The non-trivial cocycle action behave very differently,
showing its IR character which is quite distinguishable from ultra-short
distance behavior in the canonical quantization approach. Finally, in the far IR the
WZ effective theory is stable to marginal deformations by the local $R^2$ term,
{\it i.e.} the {\it a priori} arbitrary coefficient,
\begin{equation}
[b'' + {2\over 3}(b + b')]_{_{IR}} = 0\,,
\end{equation}
is a stable IR fixed point under RG flow induced by the non-trivial $\Delta_4$
operator in the Gaussian effective action $\Gamma_{WZ}$ \cite{AMa}. Hence, the
coefficient of $S_{local}^{(4)}$ in (\ref{localSf}) flows to zero logarithmically
at large distances, again just as one would expect for a UV marginally relevant but IR
marginally irrelevant operator at a Gaussian fixed point.

\section{The Energy-Momentum Tensor of $\Gamma_{WZ}$ in Four Dimensions}

We consider next the conserved energy-momentum tensors corresponding to
the non-local effective actions of the two non-trivial cocycles in $d=4$.
Using the identities in Appendix A, the tensor obtained by varying
to the first ($b$ term) in $\Gamma_{WZ}$ is
\begin{equation}
{1\over \sqrt{-g}} {\delta \over \delta\bar g^{ab}}
\int\, d^4x\, \sqrt{-\bar g}\,\overline F_4\,\sigma =
2\, \overline C_{a\ b}^{\ c\ d}\overline R_{cd}\ \sigma + 4\, \overline\nabla_c
\overline\nabla_d \left(\overline C_{(a\ b)}^{\ \ c\ d}\ \sigma\right)\,.
\label{firstT}
\end{equation}
It can be verified that the same stress tensor is obtained by first varying
the $b$ term in the $d$ dimensional anomalous action and then taking the
limit $d\rightarrow 4$. Indeed,
\begin{eqnarray}
&&{1\over \sqrt{-g}} {\delta \over \delta g^{ab}}
\int\, d^d\,x\,\sqrt{-g}\, F_d \equiv\ F_{ab} =
H_{ab} - {4\over d-2}\ ^{(2)}H_{ab} + {2\over (d-1)(d-2)}\ ^{(1)}H_{ab}
\nonumber\\
&&\qquad = 2 C_{acde}C_b^{\ cde} -{g_{ab}\over 2} C_{cdef}C^{cdef} +
{4\over d-2}\, C_{a\ b}^{\ c\ d}R_{cd} + 4\nabla_c\nabla_d  C_{(a\ b)}^{\ \ c\
d}\,,  \label{btensor}
\end{eqnarray}
where the three tensors $H_{ab}$, $^{(1)}H_{ab}$ and $^{(2)}H_{ab}$ are
defined in (\ref{Hdef}) of Appendix A. The first two terms in the last expression vanish
in $d=4$ dimensions ({\it cf.} Appendix B). However, in $d \ne 4$ dimensions
this tensor is non-zero. Let us define
\begin{equation}
C_{ab} \equiv  \lim_{d\rightarrow 4} \left\{ {g_{ab}\, C_{cdef}C^{cdef} -
4 \,C_{acde}C_b^{\ cde}\over d-4}\right\}\,.
\label{Ctensor}
\end{equation}
Evidently this tensor is non-trivial since its trace,
\begin{equation}
g^{ab}\,C_{ab} = C_{cdef}C^{cdef}
\label{Ctrace}
\end{equation}
is non-vanishing. Its $\sigma$ dependence in any dimension is very simple, due to
the conformal transformation property of the Weyl tensor,
\begin{equation}
C^a_{\ bcd} = \overline C^a_{\ bcd}\,.
\end{equation}
Using (\ref{Cderiv}) of the Appendix for the $\sigma$ dependence of the remaining
terms in (\ref{btensor}), we find
\begin{eqnarray}
&&e^{(d-2)\sigma}\, F_{ab} - \overline F_{ab} =\nonumber\\
&&\qquad 4(d-4)\left[ \overline C_{a\ b}^{\ c\ d}\sigma_{;cd} + 2 \sigma_d
\overline\nabla_c  \overline C_{a\ b}^{\ c\ d} + (d-4) \overline C_{a\ b}^{\ c\ d}
\sigma_c\sigma_d + {1\over 2}\sigma \overline C_{a\ b}^{\ c\ d}\overline R_{cd}
+  \sigma\overline\nabla_c\overline\nabla_d \overline C_{(a\ b)}^{\ \ c\ d}\right]
+ {\cal O}(d-4)^2\,.
\label{Fsvar}
\end{eqnarray}
The factor of $e^{(d-2)\sigma}$ in the first term is a consequence of the fact
that $\sqrt{-g}\, \delta g^{ab} = e^{(d-2)\sigma}\sqrt{-\bar g}\, \delta \bar g^{ab}$
in the variation holding $\sigma$ fixed, and the variation of the $C_{ab}$ term
vanishes to linear order in $d-4$. Dividing the difference of tensors (\ref{Fsvar})
by $d-4$ and taking the limit $d \rightarrow 4$ reproduces (\ref{firstT}).

Since (\ref{firstT}) is linear in $\sigma$, substituting the solution for
$\sigma$ ({\ref{solfour}),
\begin{equation}
\sigma = {1\over 2} (\varphi - \bar\varphi)\,,
\end{equation}
with the previous definition of the auxiliary field $\varphi$, (\ref{phifour})
and $\bar\varphi$ defined similarly with all terms evaluated at the
metric $\bar g_{ab}$ into (\ref{firstT}) yields the difference,
\begin{equation}
\left[C_{a\ b}^{\ c\ d} R_{cd}\,\varphi
 + 2\,\nabla_c\nabla_d \left(C_{(a\ b)}^{\ \ c\ d}\varphi\right)\right] -
\left[\overline C_{a\ b}^{\ c\ d}\overline R_{cd}\bar \varphi + 2\,\overline\nabla_c
\overline\nabla_d\left(\overline C_{(a\ b)}^{\ \ c\ d}\bar\varphi\right)\right]\,.
\end{equation}
Taking into account the $\sigma$-independent contribution from the first
two terms in (\ref{btensor}) we see that the energy-momentum tensor
obtained by varying the first cocycle ($b$ term) in the non-local anomalous
action (\ref{anomact}) is
\begin{equation}
^{(F)}T_{ab}[g]= -{2\over\sqrt{-g}}{\delta \over \delta g^{ab}}\,S_{anom}^{(F)}[g]
= b\,C_{ab}-2b\left[C_{a\ b}^{\ c\ d} R_{cd}\,\varphi
 + 2\,\nabla_c\nabla_d \left(C_{(a\ b)}^{\ \ c\ d}\,\varphi\right)\right]\,,
\label{Fstress}
\end{equation}
in $d=4$ dimensions, where now all indices are raised and lowered with
the single metric $g_{ab}$, and the correct normalization has been restored.
An explicit form for the non-local tensor $C_{ab}$ appearing in this expression
and defined by (\ref{Ctensor}) cannot be obtained from the $\sigma$ dependence
of the WZ action, but requires varying the non-local action $S_{anom}^{(F)}[g]$
directly. Comparison of (\ref{Fstress}) with (\ref{actphi}) shows that $C_{ab}$
comes from the variation of the auxiliary $\varphi$ field itself.
Since the second term in (\ref{Fstress}) is traceless, the trace
$g^{ab}\ ^{(F)}T_{ab}[g] = b C_{cdef}C^{cdef}$ comes from the $C_{ab}$ term
alone. An explicit form for $C_{ab}$ is given in Appendix C.

For the second cocycle the direct variation of the four dimensional WZ
action was calculated in ref. \cite{states}. To check this result we may evaluate
\begin{eqnarray}
{1\over \sqrt{-g}} {\delta \over \delta g^{ab}}
&&\int\, d^d\,x\,\sqrt{- g}\, E_d \equiv E_{ab} =
H_{ab} - 4\, ^{(2)}H_{ab} + ^{(1)}H_{ab}
\nonumber\\
&&\qquad = 2 C_{acde}C_b^{\ cde} - {g_{ab}\over 2}  C_{cdef}C^{cdef} + (d-4)
\left[ ^{(3)}H_{ab} + {1\over 18}\,^{(1)}H_{ab}\right]\,,
\label{secondT}
\end{eqnarray}
where
\begin{eqnarray}
^{(3)}H_{ab} &\equiv& - {4\over d-2} C_{acbd}R^{cd} + {2 (d-3)\over (d-2)^2}
R_{cd}R^{cd} g_{ab} - {4(d-3)\over (d-2)^2} R_a^{\ c}R_{bc}
\nonumber\\
&&\qquad + {2d(d-3)\over (d-1)(d-2)^2} RR_{ab} -
{(d+2)(d-3)\over 2(d-1)(d-2)^2} R^2 g_{ab}
\label{Hthree}
\end{eqnarray}
is a generalization of the tensor $^{(3)}H_{ab}$ in four dimensional conformally
flat spacetimes \cite{Bunch}. Because $E_4$ is the Euler density, the tensor $E_{ab}$
vanishes identically in $d=4$, just as the Einstein tensor $G_{ab}$ does in
$d=2$.

The same combination of Weyl squared tensors appears in (\ref{secondT})
as in (\ref{btensor}), with its simple $e^{-2\sigma}$ dependence on $\sigma$.
Considering the explicit factor of $d-4$ in the remaining terms of
(\ref{secondT}), the stress tensor corresponding to the second cocycle of the
WZ action in $d=4$ dimensions is
\begin{eqnarray}
&& ^{(E)}T_{ab}[g]= -{2\over\sqrt{-g}}{\delta \over \delta g^{ab}}\,S_{anom}^{(E)}[g]
= b'C_{ab}-2b'\left[^{(3)}H_{ab} +{1\over 18}\, ^{(1)}H_{ab}\right]_{d=4}
\nonumber\\
&& = b'\left[ C_{ab} + 4 C_{acbd}R^{cd} + {2\over 9}
\left(\nabla_a\nabla_b R - g_{ab} \sq R - 7 R R_{ab}\right)
+ {5\over 9} g_{ab} R^2 + 2R_a^{\ c}R_{bc} - g_{ab} R^{cd}R_{cd} \right]\,.
\end{eqnarray}
The conformal variation of this tensor is obtained by computing
\begin{eqnarray}
&&e^{2\sigma}\left[^{(3)}H_{ab} +{1\over 18}\,^{(1)}H_{ab}\right] -
\left[^{(3)}\bar H_{ab} +{1\over 18}\,^{(1)}\bar H_{ab}\right] =
4\bar C_{a\ b}^{\ c\ d}\left(\sigma_{;cd} - \sigma_c\sigma_d\right)
+ 4 \overline R^c_{\ (a}\sigma_{;b)c} - 4 \overline R^c_{\ (a}\sigma_{b)}\sigma_c
- {8\over 3} \overline R_{ab} \sqb \sigma
\nonumber\\
&& \qquad - {2\over 3} \overline R_{ab} \sigma_c \sigma^c
+ {2\over 3} \overline R_{;(a} \sigma_{b)} - {4\over 3} \overline R \sigma_{;ab}
+ {2\over 3} \overline R \sigma_a\sigma_b  - 4 \sigma_{;a}^{\ c}\sigma_{;bc}
+ 4 \sigma_{;ab} \sqb \sigma + {2\over 3} (\sqb \sigma)_{;ab}
+ {2\over 3} (\sigma_c\sigma^c)_{;ab} - 4 \sigma_{(a}(\sqb \sigma)_{;b)}
\nonumber\\
&&\qquad \qquad + \bar g_{ab} \left[ - 2 \overline R^{cd} \sigma_{;cd} +
{2\over 3} \overline R^{cd} \sigma_c\sigma_d - {1\over 3} \overline R_{;c}\sigma^c
+ {4\over 3} \overline R \sqb \sigma - {2\over 3} \sqb^2\sigma +
{2\over 3}\sigma_{;cd}\sigma^{;cd} - (\sqb \sigma)^2
+ {2\over 3} (\sqb \sigma)_{;c}\sigma^c\right]
\label{ourtensor}
\end{eqnarray}
A straightforward exercise in commuting covariant derivatives and use of the
Bianchi identities shows that this tensor is precisely equal to the negative of
that given by the right hand side of eq. (2.9) of ref. \cite{states} (where the notation
for the background metric $\bar g_{ab}$ omitted the overbar). In ref. \cite{states}
this energy-momentum tensor was use to study the physical state Hilbert
space of the quantized $\sigma$ field in the Einstein space background,
$R \times S^3$.

Because of (\ref{EFrel}) and (\ref{Ctrace}) the trace of the tensor $^{(E)}T_{ab}[g]$ is
\begin{equation}
g^{ab}\, ^{(E)}T_{ab}[g] = b'\left[ C_{cdef}C^{cdef} -2 R_{cd}R^{cd} +
{2\over 3} R^2 - {2\over 3}\sq R \right] = b'\left(E_4 - {2\over 3}\sq R\right)
\end{equation}
in $d=4$ dimensions.

Both the tensors $^{(F)}T_{ab}[g]$ and $^{(E)}T_{ab}[g]$ obtained
by varying the non-local anomalous action in $4$ dimensions are
themselves non-local in general. However, if the spacetime is
conformally flat then the Weyl tensor, $C_{ab}$ and
$^{(F)}T_{ab}[g]$ both vanish, while $^{(E)}T_{ab}[g]$ becomes
proportional to the local tensor $^{(3)}H_{ab} + {1\over 18}
^{(1)}H_{ab}$. Since $^{(1)}H_{ab}$ is the variation of the local
$R^2$ action, it is conserved on its own, and we conclude that
$^{(3)}H_{ab}$ with $C_{abcd} = 0$ must be conserved in $d=4$
conformally flat spacetimes. The conservation of this tensor in
this case had been noted some time ago, \cite{Bir,GinKir} and it
has been calledaccidentally conserved.' Evidently this local
tensor which is conserved in conformally flat spacetimes owes its
existence to the non-local anomalous action corresponding to the
second cocycle of the Weyl group in four dimensions. The necessary
appearance of $^{(3)}H_{ab}$ in the renormalized stress tensor of
a matter field in curved space after renormalization is explained
by the contribution of $S_{anom}$ to the effective action with a
finite coefficient, which as we have seen appears automatically in
dimensional regularization \cite{Bunch}. Because the action
$S_{anom}$ is non-invariant under global Weyl rescalings,
$^{(3)}H_{ab}$ carries information about the global scaling
behavior of a quantum theory in a state having this geometrical
term as the expectation value of its energy-momentum tensor. This
fact has been used to extract the IR $b'$ coefficient of a
massless scalar field for any curvature coupling $\xi$ by
examining the attractor behavior of its energy-momentum tensor at
late times in de Sitter space, showing that $S_{anom}$ appears
also in the effective action of theories which are not classically
Weyl invariant \cite{And}.

\section{Conformal Infinity in the FG Construction}

Since conformal invariants are central to obtaining the WZ effective action
of the non-trivial cocycles of the Weyl group by the dimensional
continuation method, we review in this section the FG procedure for constructing
such conformal invariants in any dimension. Dimensional continuation is also
the natural technique to handle a certain obstruction in the FG expansion at even
integer dimensions \cite{FG}. As discussed in the introduction, the FG embedding
may be used to classify all the IR relevant terms in the effective action on the
physical boundary space as well. Thus the treatment and applications of the FG
construction in this paper will be somewhat different than that in the recent
literature \cite{adscft,SchThe,ManMkr}.

The FG approach is based upon generalization of the following elementary
example of embedding of the round sphere $S^d$ in a flat Minkowski space of two
dimensions higher. In this example the $d+2$ dimensional space (called the ambient space)
has signature $(d+1,1)$ and flat metric $\eta_{_{AB}}$, {\it i.e.} its line element
is
\begin{equation}
d\tilde s^2 = \eta_{AB}\,dX^A\,dX^B
= -(dX^0)^2 + (dX^1)^2 + \dots + (dX^{d+1})^2
\equiv -(dX^0)^2 + d\vec X\cdot d\vec X\,,
\label{Mink}
\end{equation}
with the proper Lorentz isometry group $SO(d+1,1)$,
\begin{equation}
X^{\prime A} = \Lambda^A_{\ \ B}X^B\,.
\label{lorentz}
\end{equation}
This group has three classes of orbits in the ambient space, namely
\begin{enumerate}
\item Degenerate: The future (or past) light cone, $\eta_{AB}\,X^A\,X^B= 0$,
with $X^0 > 0$ (or $X^0 < 0$);
\item \L obachewski (Euclidean AdS): The upper (or lower) sheet of the
two-sheeted hyperboloid, $\eta_{AB}\,X^A\,X^B = -\ell^2$, with $X^0 > 0$
(or $X^0 < 0$) and negative intrinsic scalar curvature, $R = -d(d+1)/\ell^2$;
\item de Sitter: The single sheeted hyperboloid, $\eta_{AB}\,X^A\,X^B= +
\ell^2$, with positive intrinsic scalar curvature, $R= d(d+1)/\ell^2$.
\end{enumerate}
In the first class of orbits the Lorentz group acts projectively, and the
space of null directions on the future light cone is isomorphic to the sphere
$S^d$. Indeed let
\begin{equation}
N^A \equiv {X^A\over X^0} = (1, n^i)\,,\qquad i = 1, \dots d+1\,.
\end{equation}
Then $\eta_{AB}\,N^A\,N^B = 0 = -1 + n^in^i$ implies the vector $\vec n$ lies
on the unit sphere $S^d$. Under the Lorentz transformation (\ref{lorentz}),
\begin{equation}
n^i \rightarrow n^{\prime i} = {X^{\prime i}\over X^{\prime 0}} =
{\Lambda^i_{\ 0} + \Lambda^i_{\ j}\, n^j\over \Lambda^0_{\ 0} +
\Lambda^0_{\ k}\,n^k}\,. \label{proj}
\end{equation}
Using $dX^i = X^0 dn^i + n^i dX^0$, the ambient metric (\ref{Mink}) restricted
to the light cone is
\begin{equation}
ds^2_0 = (X^0)^2\, d\vec n\cdot d\vec n\,.
\label{lightc}
\end{equation}
Since the light cone is invariant under $SO(d+1,1)$ proper Lorentz transformations
in the ambient spacetime, the projective transformation (\ref{proj})
may be viewed as a local conformal transformation,
\begin{equation}
X^0\rightarrow X^{\prime\,0} = \Omega(\vec n)X^0\,,\qquad \Omega = e^{\sigma}=
\Lambda_{\ 0}^0 + \Lambda^0_{\ i}\,n^i
\end{equation}
of $S^d$. Hence the conformal structure on the sphere $S^d$ may be
investigated by studying the simpler Lorentz group in the ambient
space.

On the second class of orbits, \L obachewski space, one may introduce the
standard hyperbolic projective coordinates,
\begin{equation}
X^0 = \ell\, \left[{1 + y^2\over 1 - y^2}\right]\,,\qquad
X^i = \ell\, {2y^i\over 1-y^2}\,,\qquad y^2 \equiv \vec y\cdot\vec y\,,
\end{equation}
so that the metric takes the standard \L obachewski form,
\begin{equation}
ds^2_{_{\L}} = {4 \ell^2 d\vec y\cdot d\vec y \over (1-y^2)^2}\,,
\end{equation}
with the boundary at $y^2 = 1$. Alternately, one can introduce the coordinates
\begin{equation}
\rho = {4\over \ell^2} {(1-y)^2\over (1 + y)^2}\,, \qquad \vec n = {\vec y
\over y}
\end{equation}
to bring the \L obachewski line element into the FG form,
\begin{equation}
ds^2_{\L} = {\ell^2\over 4} \left({d\rho\over \rho}\right)^2 +
{1\over \rho }\left[\left(1- {\ell^2\rho\over 4}\right)^2 d\vec n\cdot d\vec n
\right] = {\ell^2\over 4} {d\rho^2\over \rho^2} + {1\over \rho}\,
g_{ab}(x,\rho)\,dx^a dx^b\,,
\label{FGL}
\end{equation}
with $g_{ab}(x, 0)$ the round sphere metric on $S^d$, and $x^a,\, a= 1,\dots
d$ coordinates on $S^d$. Thus the boundary of \L obachewski space at $\rho =0$
is isomorphic to the light cone of the first orbit which it approaches
asymptotically. This is the conformal boundary of $d+1$ dimensional
\L obachewski space, and in this elementary example of embedding the conformally
flat $S^d$, the \L obachewski metric (called the bulk metric)
$g_{ab}(x, \rho)$ possesses a regular Taylor expansion in $\rho$ which
terminates at the second order, after the factor of $\rho^{-1}$ has been
extracted as in (\ref{FGL}). The usefulness of this coordinatization is that
the conformal group of the boundary $S^d$ metric, $SO(d+1, 1)$ at $\rho =0$ has been
represented as the isometry group of the bulk (and ambient) space, which makes possible
the study of the conformal structure on the boundary by ordinary geometric properties
both in the $d+1$ bulk \L obachewski space, as well as the $d+2$ ambient flat space.
The Lorentzian signature AdS metric with isometry group $SO(d, 2)$ is obtained by
changing the signature of one of the $X^i$ in the ambient Minkowski space metric
(\ref{Mink}).

The analogous construction works equally well in the case of the third class of
orbits, namely de Sitter space. Indeed introducing the change of variables,
\begin{equation}
\rho = {4\over \ell^2} \exp \left(-{2\tau\over \ell}\right)
\label{rhodes}
\end{equation}
into a standard form of the de Sitter line element,
\begin{equation}
ds^2_{deS} = -d\tau^2 + \ell^2 {\rm cosh}^2\left({\tau\over \ell}\right)
d\vec n\cdot d\vec n \,,
\end{equation}
with closed $S^d$ spatial sections brings it into the FG form,
\begin{equation}
ds^2_{deS} = -{\ell^2\over 4} \left({d\rho\over \rho}\right)^2 +
{1\over \rho }\left[\left(1+ {\ell^2\rho\over 4}\right)^2 d\vec n\cdot d\vec n
\right] = -{\ell^2\over 4} {d\rho^2\over \rho^2} + {1\over \rho}\,
g_{ab}(x,\rho)\,dx^a dx^b\,,
\label{FGdeS}
\end{equation}
which is exactly the same as the Euclidean AdS case (\ref{FGL}) with $\ell^2 \rightarrow
-\ell^2$. The conformal infinity at $\rho =0$, $\tau =\infty$ is also asymptotic
to the same future light cone in the $d+2$ dimensional ambient space and
therefore it enjoys all the same conformal properties as the $\rho =0$ boundary
of Euclidean AdS. Thus, the FG method of extracting conformal invariants and
conformal field theory behavior will work equally well in asymptotically
de Sitter spacetime at its spacelike future infinity at $\rho =0$, and there
is CFT behavior at the conformal infinity of bulk de Sitter space as well.

To show the conformal behavior in de Sitter space arises more explicitly,
let us use the flat spatial sections of de Sitter space,
\begin{equation}
ds^2_{deS} = -dt^2 + \ell^2 \exp\left({2t\over \ell}\right)
d\vec x\cdot d\vec x
\end{equation}
and make the change of variables $\rho = \ell^{-2} \exp \left(-{2t\over \ell}\right)$
to bring the de Sitter line element into the alternative form,
\begin{equation}
ds^2_{deS} = -{\ell^2\over 4} \left({d\rho\over \rho}\right)^2 +
{1\over \rho }\, d\vec x\cdot d\vec x
 = -{\ell^2\over 4} {d\rho^2\over \rho^2} + {1\over \rho}\,
\delta_{ab} dx^a dx^b\,.
\label{deSFG}
\end{equation}
In this case the metric takes the FG form with $g_{ab}(x, \rho) = g_{ab}(x, 0)
= \delta_{ab}$ independent of $\rho$.
Then consider the behavior of the two-point function of a scalar field with
mass $m$ in $d+1$ dimensional de Sitter spacetime, {\it viz.} \cite{CherTag}
\begin{equation}
\langle T \Phi (t, \vec x) \Phi (t', \vec x')\rangle = {\ell^{1-d}\over
(4\pi)^{d+1\over 2}} {\Gamma ({d\over 2} - \nu) \Gamma ({d\over 2} + \nu)
\over \Gamma ({d+1\over 2})}\  _2F_1\left({d\over 2} - \nu , {d\over 2} + \nu;
{d+1\over 2}; 1- s^2(t, \vec x; t',\vec x')\right)\,,
\end{equation}
where
\begin{eqnarray}
\nu&\equiv& \sqrt {{d^2\over 4} - {m^2\ell^2}}\quad {\rm\quad
and}\nonumber\\
s^2(t,\vec x; t' \vec x') &= &{1\over 4} \exp \left( {t +
t'\over \ell}\right) \left[-\left(e^{ -{t \over \ell}} - e^{ -{t '\over
\ell}}\right)^2 + (\vec x - \vec x')^2\right]
\end{eqnarray}
depends only on the invariant distance between the two points $(t, \vec x)$ and
$(t', \vec x')$. Going to the conformal boundary at $t \sim t' \rightarrow
\infty$, $\rho\sim \rho' \rightarrow 0$, $s^2$ becomes proportional to the
flat section invariant distance $(\vec x - \vec x')^2$, and the asymptotic
behavior of the hypergeometric function $_2F_1$ implies that the two-point
function of the scalar field behaves as \cite{astroph}
\begin{equation}
\langle T \Phi (t, \vec x) \Phi (t', \vec x')\rangle \rightarrow
C_- s^{-2\Delta_-} + C_+ s^{-2\Delta_+}\,,
\label{confdes}
\end{equation}
characteristic of a conformal theory in flat space with conformal weights,
\begin{equation}
\Delta_{\pm} = {d\over 2} \pm \nu = {d\over 2} \pm \sqrt {{d^2\over 4} -
{m^2\ell^2}}\,.
\end{equation}
We note that the de Sitter case propagator comes from a massive
unitary field theory and gives complementary information about the
conformal representation corresponding to $\Delta_-$ which cannot
be obtained in the AdS case without going to tachyonic $m^2
\rightarrow -m^2$. General de Sitter correlation functions which
depend on the invariant distance between two points in the de
Sitter bulk will have conformal behavior as $\rho\rightarrow 0$ as
well, for purely geometric reasons of the embedding. The
asymptotic behavior of quantum field theory on de Sitter spacetime,
$deS_{d+1}$ induces conformal field theory behavior, $CFT_d$ on its 
asymptotic conformal infinity. This is so
because of the pure kinematic fact of group isomorphism: the de
Sitter isometry group $SO(d+1,1)$ is the same as the conformal
group $C(S^d)$ of the asymptotic conformal infinity, which
consists of the asymptotic past and the asymptotic future $S^d$ of
$deS_{d+1}$. In other words, there exists $deS_{d+1}/CFT_d$
correspondence \cite{astroph,Strom}. There is another
aspect of the $deS_{d+1}/CFT_{d}$ correspondence, namely, for even
integer $d=2k$, conformal anomalies can be computed
from the bulk gravitational theory on a spacetime of constant positive
curvature of one higher dimension. This connection with conformal
anomalies is described in the next section.

To proceed now consider the FG generalization of the above simple
example of embedding $S^d$ to an arbitrary $d$ dimensional metric.
FG showed that any $d$ dimensional metric (with the appropriate
signature) may be embedded at the conformal boundary of a space of
asymptotically constant curvature, with $g_{ab}(x, 0)= g^{(0)}_{ab}$ the
boundary metric and the bulk metric in the vicinity of the
boundary given by expanding $g_{ab}(x, \rho)$ in a Taylor series
in integer power of $\rho$, {\it i.e.}
\begin{equation}
g_{ab}(x, \rho) =  \sum_n g_{ab}^{(n)}(x) \rho^n\,.
\label{FGseries}
\end{equation}
The coefficients $g_{ab}^{(n)}$ are determined order by order in $\rho$
by solving Einstein's equations for the $d+1$ dimensional bulk metric.
However as noted by FG themselves \cite{FG} if $d=2k$ is an even integer this Taylor
series breaks down at order $n = k$, and logarithmic terms appear for the general
embedded metric. In that case only the trace and the covariant divergence of
$g_{ab}^{(k)}$ is determined by Einstein's equations in the bulk. The
remaining part of $g_{ab}^{(k)}$ cannot be determined and has been called
`the FG ambiguity' \cite{SchThe}.

The breakdown in the expansion (\ref{FGseries}) at $n = k$ for even integer
dimensions $d=2k$ has to do with the existence of certain traceless conformal
tensors on the boundary metric \cite{FG}. However, as is already apparent
from the power series solution of Einstein's equations for general $d$
in ref. \cite{HenSken}, the series (\ref{FGseries}) is well-defined for all
$n$ if $d \neq 2k$ and there are no logarithmic terms, their place being taken
instead by simple poles at $d=2n$ in the coefficients, $g_{ab}^{(n)}$. Thus
dimensional regularization supplies just the means of realizing the FG idea of
obtaining a unique $d+1$ dimensional bulk metric corresponding to an arbitrary $d$
dimensional boundary metric by solving the Einstein equations order by order in a
power series in $\rho$, with the general $g_{ab}(x,\rho)$ given by (\ref{FGseries}),
replacing the simple terminating series we found in the exact \L obachewski or de
Sitter cases. This is the first point of contact between the FG construction
and dimensional regularization.

Furthermore, the residues of the poles in $g_{ab}^{(n)}$ at $d=2n$ are linear
combinations of precisely the conserved tensors $G_{ab}$ and $F_{ab}$,
$E_{ab}$ encountered in the previous sections for $n=1$ and $n=2$
respectively, showing the close connection of these terms in the dimensionally
continued FG expansion with the non-trivial cocycles in two and four dimensions.
Indeed, the explicit form of the first few expansion coefficients in arbitrary
dimension $d$ was reported in ref. \cite{deH}, which after taking account of the
different conventions for the Riemann tensor reads
\begin{mathletters}
\begin{eqnarray}
&& g^{(1)}_{ab} = {\mp \ell^2\over d-2}\left(R_{ab} - {1\over 2(d-1)} R
g^{(0)}_{ab}\right)\,,\\
&& g^{(2)}_{ab} = {\ell^4\over d-4} \left(- {1\over 8(d-1)}
\nabla_a\nabla_b R + {1\over 4(d-2)} \sq R_{ab} - {1\over
8(d-1)(d-2)} \sq R g^{(0)}_{ab}
- {1\over 2(d-2)} R^{cd}R_{acbd}\right. \nonumber\\
&& \left.+ {d-4 \over 2 (d-2)^2} R_a^c R_{bc} +
{1\over (d-1)(d-2)^2} R\,R_{ab} + {1\over 4 (d-2)^2} R_{cd}R^{cd}\, g^{(0)}_{ab}
-{3d \over 16(d-1)^2(d-2)^2} R^2\, g^{(0)}_{ab}\right)\,.
\end{eqnarray}
\end{mathletters}
\noindent The expansion coefficients are completely well-defined and there
is no FG ambiguity in the dimensionally continued coefficients of
the expansion (\ref{FGseries}). The coefficients of the conformal
anomaly determined by this construction are the same on the de
Sitter side as on the AdS side modulo the $\ell^2\rightarrow -\ell^2$
changes in the basic formulae. The residue of the pole at $d=2$ in
$g^{(1)}_{ab}$ is proportional to the Einstein tensor, $G_{ab}$,
while the residue of the pole at $d=4$ in $g^{(2)}_{ab}$ is easily
verified by means of the identities in Appendix A to be
proportional to a linear combination of $E_{ab}$ and $F_{ab}$. In
fact, $G_{ab}$ and $E_{ab}$ vanish in two and four dimensions
respectively, and do not contribute to the pole terms, while
$F_{ab}$ is the traceless Bach tensor which is the explicit
obstruction to the $\rho$ expansion in four dimensions \cite{FG}.
The reason for this connection is that the residues of the pole
terms, which correspond to the logarithms in even integer
dimensions are necessarily conserved due to the Einstein equations
in the bulk geometry, and they transform with a definite conformal
weight $2(1-n)$ under conformal transformations of the boundary
metric. It follows from this that the residue of $g_{ab}^{(n)}$ at
$d=2n$ must be a linear combination of the conserved traceless
tensors obtained by varying the $2n$ dimensional action composed
of conformal invariants in that dimension. This is the second
point of contact between the dimensionally continued FG expansion
and the development of the previous sections.

To demonstrate the transformation properties of the coefficients $g_{ab}^{(n)}$
in detail requires the form of the coordinate transformation of the bulk
metric which leaves the FG form,
\begin{equation}
ds^2 = \pm {\ell^2\over 4} {d\rho^2\over \rho^2} + {1\over \rho}\,
g_{ab}(x,\rho)\,dx^a dx^b\,,
\label{FGform}
\end{equation}
invariant \cite{PenRin,BroHen,ISTY}. These special (PBH) transformations take the
infinitesimal
form,
\begin{eqnarray}  \rho &=& \rho' e^{-2\sigma(x')}
\simeq \rho'\,(1-2\sigma(x'))\,,\nonumber\\ x^a &=& x^{\prime\, a} +
\xi^a(x',\rho')\,.
\label{PBH}
\end{eqnarray}
Requiring the FG bulk line element have no mixed $dx^{\prime a} d\rho'$ terms
gives
\begin{equation} \xi^a(x,\rho) = \pm {\ell^2\over 2}
\int_0^{\rho}\,d\rho' g^{ab}(x, \rho') \partial_b \sigma (x) +
\xi^{a\,(0)}(x)\,.
\label{xirho}
\end{equation}
Then $\xi^a$ may be developed as a power series in $\rho$, {\it i.e.}
\begin{equation}
\xi^a(x,\rho) =
\sum_{n=0} \xi^{a\,(n)}(x)\rho^n\,,
\label{expxi}
\end{equation}
which is completely determined by the expansion (\ref{FGseries}) when we
impose the Einstein equations in the bulk geometry.
The diffeomorphism (\ref{PBH}) generates the transformation,
\begin{eqnarray}
\delta g_{ab}(x, \rho) &=& 2 \sigma (x) (1 - \rho\partial_{\rho})
g_{ab}(x,\rho) + \nabla_a \xi_b (x, \rho) + \nabla_b \xi_a(x, \rho)\qquad {\rm or}\\
\delta g_{ab}^{(n)}(x) &=& 2 (1 - n)\sigma (x)
g_{ab}^{(n)}(x) + \nabla_a \xi_b^{(n)} (x) + \nabla_b \xi_a^{(n)}(x)
\end{eqnarray}
on the metric, where $\nabla_a$ is the covariant derivative with respect to the
zeroth order boundary metric $g^{(0)}$. At $\rho = 0$, or for $n=0$ this transformation
reduces to an infinitesimal Weyl transformation (\ref{weyl}) of the boundary metric
$g_{ab}(x, \rho =0)$, up to the diffeomorphism $\xi^{(0)}_a(x)$. The transformation of
the coefficient $g_{ab}^{(n)}$ under global Weyl transformations, $\sigma (x) = \sigma_0
=const.$  is that of a tensor of weight $2(1-n)$, as required. Hence a similar
relationship of the higher order terms in the expansion, possessing poles at larger even
integer dimensions, with the variations of the Weyl invariant terms that give rise to
traceless conserved tensors in those dimensions is to be expected. Particular
linear combinations of these tensors enter the expansion (\ref{FGseries}) because the
solution of the bulk metric is determined by the second order Einstein equations on the
$d+1$ dimensional embedding space, and this solution contains no arbitrary coefficients
in non-integer dimensions.

It is important to recognize that the use of the Einstein
equations to determine the coefficients of the power series in
(\ref{FGseries}) for the bulk metric is the simplest route to
generalizing the example of the embedding of the sphere in \L
obachewsky or de Sitter space. The Einstein equations have no
dynamical content as equations of motion following from some
variational principle in this purely mathematical construction of
the bulk embedding geometry. Rather, once the bulk metric
embedding has been determined by solving the Einstein equations
for fixed constant Ricci scalar as a power series in $\rho$, we
are free to evaluate {\it any} coordinate invariant scalar action
functional on this bulk metric, and indeed this was the method FG
proposed to construct conformal Weyl invariants of the boundary
metric. Since a particular subset of bulk coordinate
transformations, the PBH transformations (\ref{PBH}) are local
conformal transformations on the boundary, coordinate invariant
scalars in the bulk give rise to conformal invariant scalars on
the boundary. In the original paper \cite{FG} the construction of
conformal invariants was proposed by considering coordinate
invariant scalars in the $d+2$ dimensional (now non-flat) ambient
space, of dimension less than $2n$ to avoid the logarithmic
obstruction to the series expansion (\ref{FGseries}). However, as
the considerations of this section show, dimensional continuation
allows one to lift this restriction, and to use invariant scalars
in the $d+1$ dimensional bulk embedding space, which contain the
same geometric information as the ambient space embedding. Thus,
dimensional continuation furnishes exactly the missing ingredient
in the FG construction, which repays the favor by furnishing the
Weyl invariants needed to construct the non-trivial cocycles in
the general procedure of the previous sections.

Moreover, as we have seen, for the general embedded geometry the region near
$\rho =0$ is similar to the light cone of the simple prototype example
of embedding $S^d$, and the expansion in powers of $\rho$ works equally well
for the metric of the $d+1$ dimensional asymptotic AdS or deS space in the vicinity
of $\rho =0$, since both spaces of positive or negative scalar curvature
asymptotically approach the same light cone, which is the conformal infinity of the $d+2$
dimensional ambient spacetime.

\section{Finite Volume Scaling and the IR Effective Action for Gravity}

The discussion in the previous section of the embedding procedure and
dimensional regularization to realize the original FG idea for constructing
the conformal invariants of the boundary metric is essentially mathematical
and kinematic in nature. The FG construction of Weyl invariants is just what is needed to
generalize the method of obtaining the anomalous WZ effective action
corresponding to non-trivial cocycles of the Weyl group in any even dimension,
as proposed in Section 2. In the physics literature to date the FG embedding has been
used almost entirely to check features of the AdS/CFT correspondence, which specifies
that the classical supergravity bulk action be used \cite{adscft}. However, the
embedding of an arbitrary metric in a space of one dimension higher, in such a way
that conformal transformations on the boundary become coordinate transformations in
the bulk has broader consequences for anomalies and infrared scaling behavior than
perhaps is evident at first. It is this physical application of the FG embedding to the
construction of the low energy effective action for gravity that we explore in this
section.

Let us consider an arbitrary coordinate invariant local action
in the $d+1$ dimensional bulk,
\begin{equation}
S_{bulk} = \int\,d\rho\,d^d\,x\,\rho^{-{d\over 2}-1}
\sqrt{g}\,B(x,\rho) \equiv \int\,d\rho\,d^d\,x\,\rho^{-{d\over 2}-1}
\sqrt{g^{(0)}}\,b(x,\rho)\,,
\label{bulkact}
\end{equation}
where we have chosen to place the $\rho$ dependence of the volume element of the
metric $g_{ab}(x, \rho)$ into the density $b(x, \rho)$.
{}From the fact that $B= \sqrt{g^{(0)}} b/\sqrt{g}$ in (\ref{bulkact})
transforms as a scalar under the PBH diffeomorphisms (\ref{PBH}), one can find the
transformation rule for $b$, namely \cite{ISTY}
\begin{equation}
\delta b(x,\rho) = -2 \sigma(x)\,\rho \partial_{\rho}b(x,\rho) +
\nabla_a\left(b(x,\rho)\xi^a(x,\rho)\right)\,,
\end{equation}
{}From this the transformation for the coefficients in the $\rho$ expansion,
\begin{equation}
b(x, \rho) = \sum_{n=0} b_n \rho^n \label{bexp}
\end{equation}
may be found, {\it i.e.}
\begin{equation}
\delta b_n = -2n\sigma b_n + \nabla_a\left(
\sum_{j=0}^{n-1} b_j\xi^{a\,(n-j)}\right)\,.
\label{dbweyl}
\end{equation}
where the $\xi^{a\, (j)}$ are the expansion coefficients of $\xi^a$
determined by (\ref{FGseries}) and (\ref{xirho}).
The form of the transformation (\ref{dbweyl}) shows that
\begin{equation}
\int \,\sqrt {g^{(0)}}\,d^{2k}x\,b_{2k}(x)
\label{binv}
\end{equation}
is invariant under local Weyl transformations in $d=2k$ dimensions, for {\it any}
diffeomorphism invariant action function $S_{bulk}$ in the bulk. Hence substituting
explicit forms for scalar functions in the bulk and carrying out the expansion of $b$
in $\rho$ to order $n=k$ in dimensionally continued $d$ dimensions by solving the
Einstein equations (with either positive or negative cosmological constant) generates
precisely the conformally invariant scalars one needs to construct the non-trivial
cocycles in $d=2k$ even dimensions. This is the explicit proof of the construction of
conformal invariants in the coordinates (\ref{FGform}). Notice also that both the
strictly local Weyl invariants (type B) for which the inhomogeneous total derivative term
in (\ref{dbweyl}) is absent, and the topological invariants (type A) for which
this term is present are both contained in the solutions of (\ref{dbweyl}).

It is instructive to write down the infinitesimal PBH conditions
which the first few $b_n$ must satisfy:
\begin{eqnarray}
\delta b_0 &=& 0\,,\nonumber\\
\delta b_1 &=& - 2\sigma b_1 \pm {\ell^2\over 2} b_0 \sq \sigma\,,\nonumber\\
\delta b_2 &=& -4\sigma b_2 \mp {b_0 \ell^2\over 4 (d-2)}
\left[R^{ab}\nabla_a\nabla_b\sigma - {1\over 2} R \sq \sigma\right]\,,
\label{varb}
\end{eqnarray}
which are solved by \cite{ISTY}
\begin{eqnarray}
b_0 &=& const.\,,\nonumber\\
b_1 &=& \pm b_0\,{\ell^2\over 4(d-1)} R\,,\nonumber\\
b_2 &=& b_0\,{\ell^4\over 32(d-2)(d-3)}\,E_4 + c_2\,\ell^4\, C_{abcd}C^{abcd}\,.
\label{bsoln}
\end{eqnarray}
The de Sitter case is again recovered from the AdS case by a
simple change of $l^2\rightarrow -l^2$ in all AdS formulae. The
coefficients $b_n$ are the Euler (type A) and Weyl (type B)
invariants which we have used in our previous construction of the
non-trivial cocycles in $2$ and $4$ dimensions. Note that the
Euler invariant in (\ref{bsoln}) is associated with the
inhomogeneous solution to (\ref{dbweyl}) induced from the lower
order $b_n$ in the second total derivative term of (\ref{dbweyl})
or (\ref{varb}), corresponding to the fact that the Euler density
is Weyl invariant only up to a total derivative. On the other hand
the local Weyl invariant $C_{abcd}C^{abcd}$ is a solution to the
homogeneous eq. $\delta b_2 = -4\sigma b_2$. Notice also that {\it
only} the non-trivial cocycles are generated by this procedure.
The trivial $R^2$ cocycle is {\it not} a solution of the PBH
equations (\ref{varb}). Clearly this is because the trivial
cocycles are not locally Weyl invariant, and only local Weyl
invariants (up to surface terms) can be generated by bulk
coordinate invariants due to the PBH symmetry. Certainly no term
like (\ref{wrongC}) is generated.

Because of the PBH symmetry $b$ necessarily satisfies the
infinitesimal form of the WZ consistency condition,
\begin{equation}
\int\,d^{2k}x\,\sqrt{g^{(0)}}(\sigma_1\delta_{\sigma_2}b -
\sigma_2\delta_{\sigma_1} b) = 0\,,
\end{equation}
since this is just a subset of diffeomorphisms of the bulk action. Since the
$b_n$ coefficients appear both in the action and in the trace anomaly after
variation with respect to $\sigma$, this excludes $\sq R$ from $b_2$ in
(\ref{bsoln}) as well. Indeed comparison of the third variation in
(\ref{varb}) with eq. (\ref{varboxR}) of Appendix A shows that $\sq R$ does not
satisfy this condition. Thus the FG construction is precisely what is required to
construct the Weyl invariants (and only those invariants) which give rise to the
non-trivial cocycles in any even dimension. Taking arbitrary linear
combinations of coordinate invariant scalars in the bulk action will give rise
to arbitrary linear combinations of the conformal invariants on the boundary,
in contrast to the AdS/CFT conjecture which applies to a specific bulk action
and specific set of anomaly coefficients in the boundary theory.

The pole term in the expansion of the metric (\ref{FGseries}) at order
$n= k$ is cancelled and does not appear in the expansion of the action density
$b(x,\rho)$. However when the expansion (\ref{bexp}) is substituted in (\ref{bulkact})
the $\rho$ integral diverges at small $\rho$ in even integer dimensions, or equivalently
the $n^{th}$ order term in the expansion of the classical bulk action possesses a pole at
$d=2n$, similar to the dimensional regularization counterterms in a quantum field theory
at the boundary. However, unlike the UV regulator of quantum theory, dimensional
regularization now appears as an IR regulator of the large volume divergences of the
classical bulk action.

The fact that the limit $\rho \rightarrow 0$ is an infrared limit
is clear also from the PBH transformation (\ref{PBH}), which shows
that $\rho \rightarrow 0$ if $\sigma = \sigma_0 \rightarrow +\infty$,
with $\rho'$ fixed. The explicit form of the change of variables
(\ref{rhodes}) to bring the de Sitter metric into the FG form also shows
conformal infinity is reached by stretching all length scales to the extreme
IR limit, where the conformal behavior of the de Sitter correlation functions
becomes apparent, as in (\ref{confdes}). This stretching of
physical length scales (with some UV cutoff imposed if necessary to regulate
short distance behavior) is precisely what is contemplated in the Wilson
description of the renormalization group \cite{Wil}. Rescaling of $\rho$ in the bulk
metric is completely equivalent to a finite volume scaling in the conformal
boundary metric, and the expansion of the general bulk action in powers
of $\rho$ is just an expansion of the general boundary action in scale
dimensions of the volume.

{}From the general PBH transformation and the explicit examples we
see that $\rho$ scales like $\lambda^{-2}$ if $\lambda$ is a
physical length scale in the boundary metric. Substituting the
expansion (\ref{bexp}) in (\ref{bulkact}) and cutting off the
lower limit of the $\rho$ integral at $\rho = \lambda^{-2}$ shows
that the dimensionally regulated series is of the form,
\begin{equation}
\sum_{n=0}^{\infty} {\lambda^{d-2n}\over d-2n} \int \,d^d\,x
\sqrt{-g^{(0)}}\,b_{2n}(x) \label{bpoles}
\end{equation}
where we neglect the $\lambda$ independent terms coming from the (fixed) upper
limit of the $\rho$ integral. In the $d$ dimensional boundary theory the
$\lambda$ dependent terms in this series have precisely the form of terms in
the effective action with mass dimension $2n$, $b_{2n}(\lambda x) =
\lambda^{-2n} b_{2n}(x)$, the power of $\lambda$ classifying their
behavior under global Weyl transformations, {\it i.e.} finite volume scaling.
The terms with $n < {d\over 2}$ are strictly relevant terms at large volumes.
Conversely those with $n > {d\over 2}$ are strictly irrelevant in the infrared limit of
large volumes. The marginally relevant terms at $d=2n$  are obtained by
taking the logarithmic variation, $\lambda {\partial\over \partial \lambda}$
which cancels the pole when $d\rightarrow 2k$, yielding the finite result
(\ref{binv}), which are the non-trivial cocycles of Weyl anomaly in $2k$ even
dimensions. As we have seen explicitly in $d=4$ the trivial $R^2$ cocycle is
not included in $b_4$. Thus the FG construction selects precisely the
infrared relevant terms and only those terms in the limit $\rho \rightarrow 0$
(or $\lambda \rightarrow \infty$).

The absence of the trivial cocycle terms which are required for UV
renormalization is a manifestation of the fact that the poles in the expansion
of the dimensionally regulated bulk action (\ref{bpoles}) are {\it infrared} poles,
their formal similarity to the UV counterterms of dimensionally regulated quantum theories
notwithstanding. Taking the logarithmic variation of (\ref{bpoles}), the
physical limit $d\rightarrow 2k$ and integrating again with respect to
$d\lambda/\lambda$ gives
\begin{equation}
\Gamma_{eff}^{IR}[g^{(0)}; \lambda]= \sum_{n=0}^{k-1} {\lambda^{2k-2n}\over 2k-2n}
\int d^{2k}x\, \sqrt{g^{(0)}}\ b_n(x) + \log\lambda\int d^{2k}x\,\sqrt{g^{(0)}}\,
b_{2k}(x) +
{\cal O} (\lambda^0)
\label{effir}
\end{equation}
which are all the infrared relevant terms in the boundary effective action
which grow with either a positive power of $\lambda$ or $\log\lambda$ as
$\lambda$ grows. We see that the non-trivial cocycles of the Weyl anomaly
are the latter which are marginally relevant under finite volume rescaling.
If $\lambda$ is replaced by $e^{\sigma_0}$ then (\ref{effir}) is exactly
of the form of the WZ consistent effective action $\Gamma_{WZ}[g^{(0)};\sigma_0]$,
augmented by the relevant local terms for $n < k$ in (\ref{locsum}) to give
the total effective action of (\ref{totalS}). The Weyl invariant terms $S_{inv}$
are order $\lambda^0$ and cannot be calculated by either method, but neither
are they relevant in the low energy, long distance limit compared to the
terms kept in (\ref{effir}) or (\ref{totalS}).

Thus the FG embedding does much more than generate just the local
Weyl invariants, which was its original purpose, and which are
needed to construct the non-trivial cocycles of the anomaly in any
even integer dimension. The terms explicitly displayed in
(\ref{effir}) which diverge as $\lambda$ and the physical volume
are taken to infinity are just the IR relevant terms of the Wilson
effective action of the generally covariant boundary theory. This
provides an unambiguous definition and extension of the Wilson RG
scale transformation to low energy effective theories of gravity. From 
this Wilson effective action point of view the integral of $R^2$ is
absent in four dimensions because it is a marginally irrelevant
operator in the IR which is independent of global Weyl rescalings,
remaining volume independent in the infinite volume $\lambda
\rightarrow \infty$ limit. Arbitrary $\lambda$ independent
non-local terms in (\ref{effir}), corresponding to $S_{inv}$ in
(\ref{totalS}) are also neutral under finite volume scale
transformations and therefore are marginally irrelevant operators
in the infrared as well. Thus, we arrive at a precise formulation of
low energy effective field theory for gravity in $d$ dimensions from 
the classical FG construction in $d+1$ dimensions. Its direct proof in
a full quantum field theory setting is more difficult than this
simple classical construction and would require that one
systematically integrate out all the fluctuations between two
scales, say $\lambda$ and $2\lambda$, to show that the new effective
action is of the same form as the previous one, (\ref{effir}) with renormalized
coefficients, provided that this effective action is used to calculate
only soft processes with momenta $p \ll \lambda$. In principle, this 
Wilson-Kadanoff exact renormalization group blocking procedure can be
applied either in the continuum or on a lattice. Although this direct
analysis would be welcome, as it would give detailed information about
the RG flow for different matter or gravitational field representations
and couplings, the general classification of terms in the effective action 
according to their properties under Weyl rescaling and the FG implementation 
of this rescaling as a coordinate transformation in one dimension higher is sufficient 
to fix the general form of the IR effective action (\ref{totalS}) or (\ref{effir}).

The low energy effective action for physical four dimensional spacetime
includes in addition to the familiar cosmological term $b_0$ which scales
as the volume, $\lambda^4$ and the Einstein-Hilbert term $b_1$ which scales
as $\lambda^2$, also the non-local $S_{anom}$ corresponding to the two
non-trivial cocycles in $d=4$ which scale as $\log\lambda$. Hence these
anomalous terms from the non-trivial cocycles are not irrelevant in the
infrared, and in principle modify the Einstein theory even at low energies
and large distances. This conclusion is perhaps less surprising if one recalls
the origin of the anomaly as the effect of massless excitations which do
not decouple at arbitrarily large distances. As in the example of
the $U(1)$ chiral anomaly, the most general low energy effective Lagrangian
consistent with symmetry contains a marginally relevant WZ term which
dominates the decay, $\pi^0 \rightarrow 2\gamma$ in the low energy limit of QCD
\cite{Witt}.

\section{Conclusions}

Since several different aspects of conformal anomalies have been presented in
this paper, and some parts of these overlap with earlier work, we summarize here the main
conclusions for the benefit of the reader:
\begin{itemize}
\item The finite shift coboundary operator of the Weyl group
may be defined without the use of anti-commuting Grassmann
variables by Eqs. (\ref{kweyl}) and (\ref{finweyl}).
\item The first cohomology of the Weyl group is defined as one-forms of
the cochain which are closed but non-exact in the sense of Eqs. (\ref{close})
and (\ref{exact}), and are represented by non-local functionals of the
metric in the physical even dimension $d=2k$.
\item The non-local cocycles of the Weyl group may be constructed in
dimensional regularization by considering all the local counterterms
of dimension $2k$ near $d=2k$, and selecting those which are conformal
invariant in and only in the physical dimension.
\item The non-trivial cocycles are of two kinds, corresponding to two kinds
of conformal invariants, those invariant up to surface terms, of which there
is only one (type A, Euler density) and those which are locally conformal invariant
(type B), of which there are an increasing number in higher even
dimensions \cite{DesSch}.
\item Both kinds of non-trivial cocycles have integrals whose local
Weyl variation vanishes linearly as $d\rightarrow 2k$, and lead to
UV finite effective actions, local in terms of $\sigma$, which automatically
satisfy the WZ consistency condition. The action obtained this way is identical
to that obtained by integrating the local anomaly with respect to $\sigma$.
\item The effective actions constructed by this method are nevertheless non-trivial
due to the multi-valuedness under the global shift (\ref{shift}), which signal
sensitivity to winding about the obstruction in the space of metrics at the singular
metrics $g_{ab} =0$ and $g^{ab} = 0$.
\item Exactly this same multi-valuedness property indicates sensitivity of the non-trivial
cocycle actions to global Weyl rescalings (\ref{global}), which imply
that they correspond to marginally relevant operators in the IR.
\item The non-local but fully covariant action corresponding to each non-trivial
cocycle may be constructed explicitly by solving a linear differential equation
for the conformal transformation $\sigma$ between two members of the
conformal equivalency class $g_{ab}$ and $\bar g_{ab}$.
\item The conformal differential operator appearing in this equation for
$\sigma$ is $\sq$ in $d=2$ dimensions and $\Delta_4$ defined in
(\ref{Deldef}) in $d=4$ dimensions. In the latter case a new
uniformization conjecture analogous to the Poincare-Yamabe
conjecture for two dimensional Riemannian manifolds suggests
itself.
\item There are analogous $d^{th}$ order conformal differential operators
on scalar functions in higher even dimensions, and although this has not
been proven in all generality, it appears to be possible by simple counting
of invariants to bring the anomalous action in higher even dimensions to a Gaussian
form in $\sigma$ by a suitable admixture of the trivial cocycle terms.
This has been verified explicitly in $d=2,4,6$ dimensions, with the
$d=6$ case treated first in ref. \cite{Breg} and also in \cite{Bas}.
In all even dimensions the propagator of the conformal invariant
differential operator is logarithmic.
\item Corresponding to each non-trivial cocycle there is a conserved energy
momentum tensor which is generally non-local also in its fully covariant form.
One of these tensors becomes the local geometric tensor $^{(3)}H_{ab}$ in
$d=4$ conformally flat spacetimes, showing the true origin of this tensor.
\item In $d=2$ there are no local Weyl invariants and the non-local effective action
found by Polyakov is the single infrared relevant term in the effective action for 2D
gravity in addition to the volume cosmological term, all other possible terms in
$S_{eff}$ being strictly irrelevant in the IR.
\item In higher dimensions the non-trivial cocycles determine the effective action
up to local terms and strictly Weyl invariant terms, as in (\ref{totalS}). Although
this is considerably less information than in two dimensions, it already
determines all the IR relevant terms in the effective action in $d=4$ and
is sufficient to preclude any term of the form (\ref{wrongC}) for the Weyl
squared anomaly.
\item The Fefferman-Graham embedding of an arbitrary space at the conformal
infinity of a higher dimensional space is exactly the construction needed
to generate the local conformal invariants, and therefore non-trivial
cocycles and WZ effective actions in all higher even dimensions by the
dimensional regularization method.
\item Conversely, dimensional regularization of the FG series (\ref{FGseries})
eliminates the obstruction or ambiguity in the original FG construction,
providing a well-defined bulk metric of either positive or negative scalar
curvature (deS or AdS) in the form (\ref{FGform}).
\item Although it makes use of the classical Einstein equations in
the bulk, the FG construction is essentially kinematic, embedding
the local Weyl group of the boundary metric in the diffeomorphisms
of the embedding space, and the asymptotically de Sitter embedding
space has the same conformal behavior at infinity in the
coordinates (\ref{deSFG}) as the asymptotically AdS embedding.
Both give rise to CFT behavior at conformal infinity $\rho
\rightarrow 0$. In other words, there is
$deS_{d+1}/CFT_d$ correspondence as well.
\item By cataloging all scalar invariants in the $d+1$ dimensional
bulk geometry of the dimensionally continued FG construction, whose
volume integrals diverge as the conformal limit $\rho \rightarrow 0$
is approached, all IR relevant terms (and only those terms) in the low energy
effective action of gravity in any integer dimension may be obtained. In this case
dimensional regularization may be used as well, with poles appearing to regulate now
the IR (instead of UV) divergences at infinite volume.
\item Hence, the FG embedding of the local Weyl group into diffeomorphisms of one
higher dimension gives a precise meaning to finite volume rescaling in the Wilson RG
sense to theories possessing low energy general coordinate invariance.
\item The non-trivial cocycles of the Weyl group and the non-local effective
actions they generate are marginally relevant in the IR, while the trivial
cocycles and Weyl invariant terms in (\ref{totalS}) or (\ref{effir}) are either
marginally or strictly irrelevant ($\lambda^p, \,p\le 0$) and may be neglected in the low
energy effective action of gravity.
\item The known anomalies in $d=4$ lead to effective actions which are
marginally relevant, implying modification of the classical Einstein theory
at low energies or large distances.
\end{itemize}

Finally the higher order differential operator in $d=4$ leads to no problems
with ghosts or unphysical poles \cite{states}, and indicates instead an additional global
scalar degree of freedom in the low energy effective theory of $d=4$ gravity
over and above the Einstein theory. Since the conformal mode is completely frozen in
the classical Einstein theory, the new mode can lead to qualitatively new effects in
the modified theory. In fact, the fluctuations of this new degree of
freedom in the quadratic action (\ref{actphi}) generate an infrared
stable Gaussian fixed point characterized by restoration of conformal symmetry and
anomalous scaling of both the Einstein-Hilbert and cosmological terms \cite{AMa}.
At this fixed point the scaling dimensions of the Einstein and cosmological terms
are different from their classical dimension in the classification of relevant, marginal
and irrelevant terms in the low energy effective action of the previous section. These
refer to the perturbative (therefore also Gaussian) fixed point of flat spacetime where
scaling dimensions of terms under the RG are given by their canonical dimensions. The new
non-perturbative fixed point found in ref. \cite{AMa} describes a conformal invariant
phase of strong gravity in $d=4$ where the effective cosmological $\Lambda$ and inverse
Newtonian $G^{-1}$ terms flow to zero and it is the $S^{(E)}_{anom}$ term of
(\ref{actphi}) which controls the physics \cite{AMMb}. Possible consequences of this
conformal invariant phase of gravity for the Cosmic Microwave Background have been
investigated in \cite{AMMc}.

%\nopagebreak
%\samepage
\vspace{.75cm}
\noindent{\bf Acknowledgments}

One of the authors (P. O. M.) gratefully acknowledges the financial support of Los Alamos
Laboratory, T-8, where this work was initiated and C. N. R. S. during his sabbatical
leave of absence from the Univ. of S. Carolina. Both authors wish to thank C. N. R. S.,
the Ecol{\`e} Polytechnique, Palaiseau, France, and I. Antoniadis in particular for his
hospitality and for stimulating conversations at the outset of this work. The research of
P. O. M. was supported also by the NSF grant 9971005 to the University of South Carolina.

%\begin{thebibliography}{99}

%\end{thebibliography}

\pagebreak
\newpage
\appendix
\section{Conformal Variations}

For the reader's convenience we catalog in this Appendix the various tensors
and their conformal variations needed to derive the detailed formulae of
the text. The conformal variations may be derived from the relation between
the covariant derivative with respect to the metric $\bar g_{ab}$ and
\begin{equation}
g_{ab} = e^{2\sigma}\bar g_{ab}.
\end{equation}
By definition,
\begin{mathletters}
\begin{eqnarray}
\nabla_a V^b_{\ c} \equiv \partial_a V^b_{\ c} + \Gamma^b_{\ ad} V^d_{\ c}
- \Gamma^d_{\ ac} V^b_{\ d}\,;\\
\overline\nabla_a V^b_{\ c} \equiv \partial_a V^b_{\ c} + \overline\Gamma^b_{\ ad}
V^d_{\ c} - \overline\Gamma^d_{\ ac} V^b_{\ d}\,.
\end{eqnarray}
\end{mathletters}
\noindent where the Christoffel connections are related by
\begin{eqnarray}
\Gamma^b_{\ ac} &\equiv& {g^{bd}\over 2}\left( -\partial_d g_{ac} +\partial_a g_{cd}
+ \partial_c g_{ad}\right)\nonumber\\
&=& {\bar g^{bd}\over 2}\left( -\partial_d \bar g_{ac} +\partial_a \bar g_{cd}
+ \partial_c \bar g_{ad}\right) + \bar g^{bd}\left( -\bar g_{ac}\partial_d\sigma
  +\bar g_{cd}\partial_a \sigma + \bar g_{ad}\partial_c \sigma\right)\nonumber\\
&=& \overline \Gamma^b_{\ ac} + \Delta\Gamma^b_{\ ac}\,,
\label{covar}
\end{eqnarray}
with
\begin{equation}
\Delta\Gamma^b_{\ ac} \equiv - \bar g_{ac}\sigma^b +\delta^b_{\ c} \sigma_a +
\delta^b_{\ a} \sigma_c\,.
\label{delGam}
\end{equation}
This is the basic relation from which all the needed conformal variations may
be derived. From these relations and the definition of the Riemann tensor,
\begin{equation}
[\nabla_c, \nabla_d]\, v^a = R^a_{\ bcd}\,v^b
\end{equation}
we obtain its conformal variation,
\begin{equation}
R^a_{\ bcd} = \overline R^a_{\ bcd} + 2\delta^a_{\ [d} \sigma_{;c]b}
+ 2\bar g_{b [c}\sigma_{;d]^a}
+ 2\bar g_{b [d}\sigma_{c]}\sigma^a + 2\delta^a_{\ [c}\sigma_{d]}
\sigma_b + 2\delta^a_{\ [d}\bar g_{c]b}\sigma_e\sigma^e \,,
\end{equation}
where $\sigma_a \equiv \overline\nabla_a \sigma$ and $\sigma_{;ab} \equiv
\overline\nabla_b\overline\nabla_a\sigma = \overline\nabla_a\overline\nabla_b\sigma$,
all barred covariant derivatives taken with respect to the metric $\bar g_{ab}$,
and $2V_{[ab]} \equiv V_{ab} - V_{ba}$ denotes anti-symmetrization of
the bracketed indices. The contractions of this formula in $d$ dimensions,
\begin{mathletters}
\begin{eqnarray}
R_{cd} &=& \overline R_{cd} - (d-2)\left(\sigma_{;cd} - \sigma_c\sigma_d\right)
- \bar g_{cd} \left[\sqb \sigma + (d-2)\sigma_a\sigma^a\right]\,;\\
R &=& e^{-2\sigma} \left\{\overline R - 2(d-1)\sqb \sigma - (d-1)(d-2)\sigma_a\sigma^a
\right\}\,,
\end{eqnarray}
\end{mathletters}
\noindent follow immediately.

The conformal factor dependence of various tensors quadratic
in the curvature may be worked out next:
\begin{mathletters}
\begin{eqnarray}
R_a^{\ c}R_{bc} &=& e^{-2\sigma} \left\{ \overline R_a^{\ c}\overline R_{bc} -2(d-2)
\overline R^c_{\ (a}\sigma_{;b)c} + 2 (d-2)\overline R^c_{\ (a} \sigma_{b)}\sigma_c
-2\overline R_{ab} \left[\sqb \sigma + (d-2)\sigma_c\sigma^c\right]\right.
\nonumber \\
&& \qquad + (d-2)^2 \left(\sigma_{;a^c} - \sigma_a\sigma^c\right)
\left(\sigma_{;bc} - \sigma_b\sigma_c\right)
+ (d-2) \left(\sigma_{;ab} - \sigma_a\sigma_b\right)
\left[\sqb \sigma + (d-2)\sigma_c\sigma^c\right]
\nonumber\\
&&\qquad \qquad\left. + \bar g_{ab} \left[ (\sqb \sigma)^2 + 2 (d-2) (\sqb\sigma)\
\sigma_c\sigma^c + (d-2)^2 (\sigma_c\sigma^c)^2 \right] \right\}\,;\\
R^{cd}R_{cd} &=& e^{-4\sigma} \left\{\overline R^{cd}\overline R_{cd}
- 2(d-2) \overline R^{cd}\left(\sigma_{;cd} - \sigma_c\sigma_d\right)
- 2(d-2) \overline R \sigma_c\sigma^c - 2 \overline R \sqb \sigma
+ (3n - 4)(\sqb \sigma)^2 \right.
\nonumber\\
&&\qquad \left. + (d-2)^2 \left(\sigma_{;cd} - \sigma_c\sigma_d\right)
\left(\sigma^{;cd} - \sigma^c\sigma^d\right)
+ 2 (d-2)(2d-3) (\sqb \sigma)\sigma^c\sigma^c
+ (d-2)^3 (\sigma^c\sigma^c)^2\right\}\,;\\
RR_{ab} &=& e^{-2\sigma} \left\{ \overline R\,\overline R_{ab} - 2(d-1)\overline R_{ab}
\sqb \sigma - (d-1)(d-2)\overline R_{ab} \sigma^c\sigma^c - (d-2) \overline R
\left(\sigma_{;ab} - \sigma_a\sigma_b\right) \right.
\nonumber\\
&&\qquad - \bar g_{ab} \overline R \left[\sqb \sigma + (d-2) \sigma_c\sigma^c\right]
+ (d-1)(d-2) \left(\sigma_{;ab} - \sigma_a\sigma_b\right)
\left[2\sqb \sigma + (d-2) \sigma_c\sigma^c\right]
\nonumber\\
&& \qquad \qquad\left. + (d-1)\bar g_{ab} \left[2 (\sqb \sigma)^2 +
3 (d-2) (\sqb\sigma)\ \sigma_c\sigma^c
+ (d-2)^2 (\sigma_c\sigma^c)^2 \right] \right\}\,;\\
R^2 &=& e^{-4\sigma} \left\{\overline R^2 - 4(d-1) \overline R\ \sqb \sigma
- 2 (d-1)(d-2) \overline R \sigma_c\sigma^c + 4 (d-1)^2 (\sqb \sigma)^2\right.
\nonumber\\
&& \qquad\left.+ (d-1)^2 (d-2) (\sqb \sigma)\ \sigma_c\sigma^c + (d-1)^2(d-2)^2
(\sigma_c\sigma^c)^2\right\}\,.
\label{Rquad}
\end{eqnarray}
\end{mathletters}
\noindent From these conformal dependences we derive
\begin{eqnarray}
&&\sqrt{-g} \left(R_{ab}R^{ab} - {1\over 3}R^2\right) =
\sqrt{-\bar g} \left(\overline R_{ab}\overline R^{ab} - {1\over 3}\overline R^2\right)
\nonumber\\
&& + \sqrt{-\bar g} \left\{ -4 \overline R^{ab} (\sigma_{;ab} -\sigma_a\sigma_b)
+ 2 \overline R \sigma_{;a}^a -4 (\sigma_{;a}^a)^2 - 4 \sigma_{;a}^a \sigma_b
\sigma^b +4 \sigma_{;ab}( \sigma^{;ab} - 2 \sigma^a \sigma^b)\right\}
\nonumber \\
&&+ (d-4) \sigma\left\{R_{ab}R^{ab} - {1\over 3}R^2 -4 \overline R^{ab}
(\sigma_{;ab} -\sigma_a\sigma_b) + 2 \overline R \sigma_{;a}^a
-4 (\sigma^a_{;a})^2 - 4 \sigma_{;a}^a \sigma_b\sigma^b +
4 \sigma_{;ab}( \sigma^{;ab} - 2 \sigma^a \sigma^b)\right\}
\nonumber\\
&&+ (d-4)\Big\{ -2\overline R^{ab} (\sigma_{;ab}-\sigma_a\sigma_b)
+ {4\over 3} \overline R \sigma_{;a}^a + {4\over 3} \overline R \sigma_a\sigma^a
- 5 (\sigma^a_{;a})^2 + 4 \sigma_{;ab}( \sigma^{;ab} - 2 \sigma^a \sigma^b)
\nonumber\\ &&
-10 \sigma_{;a}^a \sigma_b\sigma^b - 4 \sigma_a\sigma^a\sigma_b\sigma^b
\Big\} + {\cal O}(d-4)^2\,.
\end{eqnarray}
The terms involving $\sigma$ but no factor of $d-4$ in the second
line of this expression can be written as a total derivative, and
hence give only a surface term when integrated. Ignoring any such
surface contributions we have then
\begin{eqnarray}
&&\int\,\sqrt{-g}\,d^d\,x \left(R_{ab}R^{ab} - {1\over 3}R^2\right) -
\int\,\sqrt{-\bar g}\,d^d\,x \left(\overline R_{ab}\overline R^{ab} - {1\over 3}
\overline R^2\right) = (d-4) \int\,\sqrt{-\bar g}\, \sigma
\left(\overline R_{ab}\overline R^{ab} - {1\over 3}\overline R^2\right)
\nonumber\\
&& - 4(d-4) \int\,\sqrt{-\bar g}\, d^d\,x\, \sigma
\left\{ -(\overline R^{ab} - {1\over 2} \bar g^{ab} \overline R)\sigma_b
\sigma^{;ab}\sigma_b - \sigma^a \sqb\sigma - \sigma^a\sigma_b\sigma^b\right\}_{;a}
\nonumber\\
&&+ (d-4)\int\,d^d\,x \sqrt{-\bar g}\, \Big\{ -2\overline R^{ab}
(\sigma_{;ab}-\sigma_a\sigma_b)
+ {4\over 3} \overline R \sigma_{;a}^a + {4\over 3} \overline R \sigma_a\sigma^a
- 5 (\sigma^a_{;a})^2 + 4 \sigma_{;ab}( \sigma^{;ab} - 2 \sigma^a \sigma^b)
\nonumber\\ &&
\qquad -10 \sigma_{;a}^a \sigma_b\sigma^b - 4 \sigma_a\sigma^a\sigma_b\sigma^b\Big\}
+ {\cal O}(d-4)^2\nonumber\\
=&& (d-4) \int\,\sqrt{-\bar g}\, \sigma  \left(\overline R_{ab}\overline R^{ab} -
{1\over 3}\overline R^2 + {1\over 3}\sqb \overline R\right) -(d-4)\int\,d^d\,x
\sqrt{-\bar g}\, \sigma\bar\Delta_4\sigma + {\cal O}(d-4)^2\,,
\label{Ricexp}
\end{eqnarray}
up to terms of linear order in $d-4$ in the expansion around $d=4$. This is the formula
used in (\ref{Rexp}) of the text.

Following standard notation we define the three tensors,
\begin{mathletters}
\begin{eqnarray}
H_{ab}&& \equiv {1\over \sqrt{-g}}{\delta\over\delta g^{ab}}\int\,d^d\,\,x\,\sqrt{-g}
R_{cdef}R^{cdef}
= 2 R_a^{\ cde}R_{bcde} - {g_{ab}\over 2} R^{cdef}R_{cdef}
+ 4\sq R_{ab} \\
&& \qquad\qquad - 2 \nabla_a\nabla_b R - 4R_a^{\ c}R_{bc} + 4 R^{cd}R_{acbd}\,;
\\
^{(2)}H_{ab}&& \equiv  {1\over \sqrt{-g}}{\delta\over\delta g^{ab}}
\int\,d^d\,\,x\,\sqrt{-g}  R_{cd}R^{cd} =
2 R^{cd}R_{acbd} - {g_{ab}\over 2} R^{cd}R_{cd}
-\nabla_a\nabla_b R + 2 \sq R_{ab} + {g_{ab}\over 2}\sq R \,;\\
^{(1)}H_{ab}&&\equiv {1\over \sqrt{-g}}{\delta\over\delta g^{ab}}
\int\,d^d\,\,x\,\sqrt{-g} R^2 =
2 g_{ab}\sq R  -2\nabla_a\nabla_b R + 2 R R_{ab} - {g_{ab}\over 2}R^2 \,.
\end{eqnarray}
\label{Hdef}
\end{mathletters}
\noindent These differ from ref. \cite{Bunch} by an overall minus sign due to
neglect of the sign change between $\delta g_{ab}$ and $\delta g^{ab}
= -g^{ac}g^{bd}\delta g_{bd}$ in ref. \cite{Bunch}.
The definition of the Weyl tensor in $d$ dimensions is
\begin{equation}
C^a_{\ bcd} \equiv R^a_{\ bcd} + {2\over d-2}\left( \delta^a_{\ [d}R_{c]b} +
g_{b[c}R_{d]}^a\right) + {2\over (d-1)(d-2)} \delta^a_{\ [c}g_{d]b}R\,,
\end{equation}
which is conformally invariant in $d$ dimensions, {\it i.e.}
\begin{equation}
C^a_{\ bcd} = \bar C^a_{\ bcd}\,.
\end{equation}
The quadratic contractions,
\begin{eqnarray}
C_a^{\ cde}C_{bcde} &=& R_a^{\ cde}R_{bcde} - {4\over d-2}R_{a\ b}^{\ c\ d}R_{cd}
+ {2\over (d-2)^2} \left( 2RR_{ab} - n R_a^cR_{bc}\right) \nonumber\\
&& + {2 \over (d-2)^2} g_{ab} R_{cd}R^{cd} - {2\over (d-1)(d-2)^2} g_{ab} R^2\,,
\end{eqnarray}
and
\begin{equation}
C_{cdef}C^{cdef} = R_{cdef}R^{cdef} - {4\over d-2}R_{cd}R^{cd}
+ {2\over (d-1)(d-2)} R^2\,,
\end{equation}
also have simple transformation properties under the local Weyl group.

{}From the transformation rule for covariant derivatives
(\ref{covar}) and {\ref{delGam}) one may derive also
\begin{eqnarray}
\nabla_a \nabla_b R &=& e^{-2\sigma} \left\{ \overline R_{;ab} - 6
\overline R_{;(a}\sigma_{b)} - 2 \overline R \sigma_{;ab} +
8 \overline R\ \sigma_a\sigma_b
- 2 (d-1)(\sqb \sigma)_{;ab} - (d-1)(d-2)(\sigma_c\sigma^c)_{;ab} \right.
\nonumber\\
&&\qquad + 4 (d-1)(\sqb \sigma)\sigma_{;ab} + 12 (d-1) (\sqb \sigma)_{;(a}\sigma_{b)}
+ 2 (d-1)(d-2)\sigma_{;ab} \sigma_c\sigma^c
\nonumber\\
&& \qquad  + 12 (d-1)(d-2) \sigma_{(a} \sigma_{;b)c}\sigma^c
-16 (d-1)\sigma_a\sigma_b\, \sqb \sigma
- 8 (d-1)(d-2)\sigma_a\sigma_b \sigma_c\sigma^c
\nonumber\\
&&\qquad\qquad + \bar g_{ab} \left[ \overline R_{;c}\sigma^c - -2
\overline R\ \sigma_c\sigma^c
-2 (d-1)(\sqb \sigma)_{;c}\sigma^c - 2 (d-1)(d-2)\sigma_{;cd}\sigma^{;cd}\right.
\nonumber\\
&&\qquad\qquad\quad \left.\left. + 4 (d-1) (\sqb \sigma)\ \sigma_c\sigma^c
+ 2(d-1)(d-2) (\sigma_c\sigma^c)^2\right]\right\} \,,
\end{eqnarray}
and
\begin{eqnarray}
\sq R &=&e^{-4\sigma} \left\{ \sqb\overline R + (d-6)\overline R_{;c}\sigma^c
- 2 (d-1) (d-2) \overline R_{cd}\sigma^c\sigma^d - 2 \overline R \sqb\sigma
- 2(d-4) \overline R \sigma_c\sigma^c - 2 (d-1) \sqb^2\sigma\right.
\nonumber\\
&&\qquad- 2 (d-1)(d-2) \sigma_{;cd}\sigma^{;cd}
 - 4(d-1)(d-4)(\sqb \sigma)_{;c}\sigma^c
- 2(d-1)(d-2)(d-6) \sigma_{;cd} \sigma^c\sigma^d
\nonumber\\
&&\qquad\qquad \left.+ 4 (d-1)(\sqb \sigma)^2 + 2 (d-1)(3d-10) (\sqb \sigma)\
\sigma_c\sigma^c + 2 (d-1)(d-2)(d-4) (\sigma_c\sigma^c)^2\right\}\,,
\label{varboxR}
\end{eqnarray}
and finally,
\begin{equation}
\nabla_c\nabla_d C_{(a\ b)}^{\ \ c\ d} = e^{-2\sigma} \left\{
\overline\nabla_c\overline\nabla_d \overline C_{(a\ b)}^{\ \ c\ d}
+ (d-3) \overline C_{(a\ b)}^{\ \ c\ d} \sigma_{;cd} + 2(d-4) \sigma_d
\overline\nabla_c \overline C_{(a\ b)}^{\ \ c\ d} + (d-3)(d-5)
\overline C_{(a\ b)}^{\ \ c\ d}\ \sigma_c\sigma_d \right\}\,.
\label{Cderiv}
\end{equation}

\section{An Identity of the Weyl Tensor in Four Dimensions}

Using the van der Waerden method the Weyl tensor may be decomposed
into its irreducible components, $(2,0)$ and $(0,2)$ of $SL(2,{\mathcal
C})\otimes SL(2,{\mathcal C})$  corresponding to the self-dual and the
anti-self-dual parts $C_{abcd}=C_{abcd}^{(+)}+C_{abcd}^{(-)}$ \cite{PenRin}.
The Weyl tensor itself corresponds to the representation $(2,0)\oplus (0,2).$
The finite-dimensional irreducible representations of $SL(2,\mathcal{C})$
are given by the space of complex completely symmetric spinors with the
number of indices equal to the twice the total spin, $2s$.
The completely symmetric spinor with $N(s)$ indices has $N(s) = 2s+1$
independent components, which is exactly the dimension of the angular
momentum $s$ representation of $SU(2)\subset SL(2,\mathcal{C})$.
The explicit mapping between $d=4$ spacetime or tangent space indices and
two-component spinorial indices is given by the Pauli-van der Waerden matrices
$\sigma_{A\stackrel{.}{B}}^a$. The decomposition of the Weyl tensor
into its self-dual and anti-self-dual components correspond in the spinorial
description to
\begin{equation}
C_{abcd}=C_{abcd}^{(+)}+C_{abcd}^{(-)} \Leftrightarrow \Psi_{ABCD
\text{ }}\epsilon_{\stackrel{.}{A}\stackrel{.}{B}}
\epsilon_{\stackrel{.}{C}\stackrel{.}{D}}+
\overline{\Psi}_{\stackrel{.}{A}\stackrel{.}{B}\stackrel{.}{C}\stackrel{.}{D}
\text{ }}\epsilon_{AB}\epsilon _{CD}\,.
\label{WeylSpinor}
\end{equation}
The anti-symmetric spinors $\epsilon _{AB}$ and
$\epsilon_{\stackrel{.}{A}\stackrel{.}{B}}$ correspond  to the `metric' on
the space of spinors; they are used to lower upper spinor  indices and their
`contravariant' form raises spinor indices. Care must be taken as
far as the order of spinor indices is concerned because of
the anti-symmetric nature of $\epsilon $'s. Each of these representations
in (\ref{WeylSpinor}) have a definite helicity and in physical terms the
linearized Weyl tensor decomposition may be identified with the spin-2
helicity $\pm 2$ wave functions of the graviton field.

With this short introduction to $SL(2,{\mathcal C})$ spinors,
we are ready to demonstrate that the symmetric tensor,
\begin{equation}
C^{acde}C_{bcde}-\frac{1}{4}\delta _b^aC^{cdef}C_{cdef}\equiv 0\,,
\label{BunchID}
\end{equation}
vanishes identically in $d=4$ dimensions.
Let us compute the first term in the above identity using the
spinorial notation, (\ref{WeylSpinor})
\begin{eqnarray}
C^{acde}C_{bcde} &\Leftrightarrow &\left( \Psi ^{ACDE}\text{ }\epsilon
^{\stackrel{.}{A}  \stackrel{.}{C}}\epsilon
^{\stackrel{.}{D}\stackrel{.}{E}}+\overline{\Psi }^{
\stackrel{.}{A}\stackrel{.}{C}\stackrel{.}{D}\stackrel{.}{E}}\text{ }%
\epsilon ^{AC}\epsilon ^{DE}\right) \left( \Psi _{BCDE}\text{ }\epsilon _{%
\stackrel{.}{B}\stackrel{.}{C}}\epsilon _{\stackrel{.}{D}\stackrel{.}{E}}+%
\overline{\Psi }_{\stackrel{.}{B}\stackrel{.}{C}\stackrel{.}{D}\stackrel{.}{E%
 }}\text{ }\epsilon _{BC}\epsilon _{DE}\right)\nonumber\\   &&= \Psi
^{ACDE}\Psi _{BCDE}\epsilon ^{\stackrel{.}{A}\stackrel{.}{C}}\epsilon
^{\stackrel{.}{D}\stackrel{.}{E}}\epsilon _{\stackrel{.}{B}\stackrel{.}{C}%
}\epsilon _{\stackrel{.}{D}\stackrel{.}{E}}+c.c=2\Psi ^{ACDE}\Psi
_{BCDE}\delta_{\stackrel{.}{B}}^{\stackrel{.}{A}}+c.c \nonumber\\
&&=\Psi ^{CDEF}\Psi _{CDEF}\delta _{\text{ }B}^{A}\delta
_{\ \stackrel{.}{B}}^{\stackrel{.}{A}}+c.c.=\left( \Psi ^{CDEF}\Psi
_{CDEF}+\overline{\Psi }%
^{\stackrel{.}{C}\stackrel{.}{D}\stackrel{.}{E}\stackrel{.}{F}}\text{ }%
\overline{\Psi }_{\stackrel{.}{C}\stackrel{.}{D}\stackrel{.}{E}\stackrel{.}{F%
 }}\right) \delta _{\text{ }B}^{A}\delta _{\ \stackrel{.}{B}}^{%
\stackrel{.}{A}}\,,
\label{psifirst}
\end{eqnarray}
where we have used the fact that the contraction over three indices of
the product of two $\Psi $'s is simply given by the formula following from
the symmetry of $\Psi $ and the antisymmetry of $\epsilon$, namely
\begin{equation}
\Psi ^{ACDE}\Psi _{BCDE}=\frac{1}{2}\Psi ^{CDEF}\Psi _{CDEF}\delta _{\text{ }%
B}^{A}\,.
\label{psiunc}
\end{equation}
On the other hand the square of the Weyl tensor is given by the contraction
over $A$, $B$ and $\stackrel{.}{A},\stackrel{.}{B}$ of
(\ref{psifirst}), {\it i.e.}
\begin{equation}
C^{bcde}C_{bcde}=4\left( \Psi ^{BCDE}\Psi _{BCDE}+\overline{\Psi }^{%
\stackrel{.}{B}\stackrel{.}{C}\stackrel{.}{D}\stackrel{.}{E}}\text{ }%
\overline{\Psi }_{\stackrel{.}{B}\stackrel{.}{C}\stackrel{.}{D}\stackrel{.}{E%
}}\right)\,.
\label{psisq}
\end{equation}
Converting $\delta _{\text{ }B}^{A}\delta _{\stackrel{.}{\ \text{ }B}}^{%
\stackrel{.}{A}}$ back to spacetime indices by multiplying (\ref{psifirst}) by
the Pauli matrices $\sigma_{A\stackrel{.}{B}}^a$ and taking account of
(\ref{psisq}) then yields (\ref{BunchID}).

\section{The Tensor $C_{ab}$}

In this Appendix we compute the nonlocal tensor $C_{ab}$ defined by (\ref{Fstress})
of the text. If we use the definition of the auxiliary field $\varphi$ given by
(\ref{phifour}) and vary the action corresponding to the $F$ cocycle of the $d=4$
anomaly given by the first term of (\ref{actphi}) we obtain
\begin{equation}
\delta S_{anom}^{(F)}[g]=\frac{b}{2} \int d^4x\,
\Biggl(\delta(\sqrt{-g}\, F_4) \varphi + \sqrt{-g}\,\, F_4
\delta\varphi \Biggr)\,.
\end{equation}
The first term in the variation of $S_{anom}^{(F)}$ is already
known and it is presented in the text. The second term leads to
$C_{ab}$, which will be computed below. We need to compute the
variation of $\varphi$ which is
\begin{equation}
\delta{\varphi(x)}=\frac{1}{2} \int d^4x' \Biggl[\delta
\left(\sqrt{-g'} (E_4 - {2\over 3} \sq R)'\right) D_4(x,x') +
\sqrt{-g'}\, (E_4 - {2\over 3} \sq R)'
\delta D_4(x,x')\Biggr]\,.
\label{varofphi}
\end{equation}
The variation of the Green's function $D_4(x,x')$ is
\begin{equation}
\delta D_4(x,x')= - \int d^4x'' D_4(x,x'')
\delta \left(\sqrt{-g}\,\Delta_4\right)'' D_4(x'',x')\,.
\label{varofD}
\end{equation}
Defining a new non-local scalar $\psi$ by the formula,
\begin{equation}
\psi(x) \equiv \frac{1}{2} \int d^4x''\, (F_4)'' D_4(x'',x)\,,
\end{equation}
we obtain the following formula for the total variation of $S_{anom}^{(F)}$:
\begin{equation}
\delta S_{anom}^{(F)}[g]=\frac{b}{2} \int d^4x\,
\Biggl[\delta(\sqrt{-g}\,\, F_4) \varphi + \sqrt{-g}\,\psi
\delta\left(E_4 - {2\over 3} \sq R\right) - 2\,\sqrt{-g}\,\psi
\delta(\Delta_4)\varphi \Biggr]\,,
\end{equation}
after taking advantage of the cancellation of terms obtained by varying
the two $\sqrt{-g}$ factors in (\ref{varofphi}) and (\ref{varofD}).
Using the relation (\ref{EFrel}) for $d=4$, we can rewrite the variation of
$S_{anom}^{(F)}$ in the form,
\begin{equation}
\delta S_{anom}^{(F)}[g]=\frac{b}{2} \int
d^4x\,\Biggl[\delta(\sqrt{-g}\,\, F_4) \varphi + \sqrt{-g}\, \psi
\delta F_4 - 2\psi\delta (\Delta_4)\varphi  -
2\psi \delta\left(R^{ab}R_{ab} - \frac{1}{3}R^2 +
\frac{1}{3}\sq R \right)\Biggr]\,.
\end{equation}
We shall evaluate all four terms in the variation $\delta
S_{anom}^{(F)} = \delta S_1^{(F)} + \delta S_2^{(F)} + \delta S_3^{(F)}
+ \delta S_4^{(F)}$, where
\begin{equation}
\delta S_1^{(F)} = \frac{b}{2} \int d^4x\,\delta (\sqrt{-g}\, F_4)\varphi\,,
\end{equation}

\begin{equation}
\delta S_2^{(F)} = \frac{b}{2} \int d^4x\,\sqrt{-g}\, \psi\, \delta{F_4}\,,
\end{equation}

\begin{equation}
\delta S_3^{(F)} = -b \int d^4x\,\sqrt{-g}\,
\psi\,\delta \Delta_4\, \varphi
\end{equation}

\begin{equation}
\delta S_4^{(F)} = -b \int d^4x\,\sqrt{-g}\,
\psi\,\delta\left(R^{ab}R_{ab} - \frac{1}{3}R^2 +
\frac{1}{3}\sq R \right)\,.
\end{equation}
Let $\delta g^{ab} = h^{ab}$, then we have

\begin{equation}
\delta S_1^{(F)} = \frac{b}{2} \int d^4x \sqrt{-g}\,
h^{ab}\Biggl(2R^{cd}C_{acbd}\varphi + 4
\nabla^c \nabla^d (C_{acbd}\varphi)\Biggr)\,,
\end{equation}

\begin{equation}
\delta S_2^{(F)} = \frac{b}{2} \int d^4x \sqrt{-g}\, h^{ab}
\Biggl(2R^{cd}C_{acbd} \psi + 4\nabla^c \nabla^d (C_{acbd}\psi)
+ \frac{1}{2} g_{ab} F_4 \psi \Biggr)\,,
\end{equation}

\begin{eqnarray}
&&\delta S_3^{(F)} = \frac{b}{2} \int d^4x \sqrt{-g}\,
h^{ab}\left[-g_{ab}\left(\psi\sq^2\varphi  - \sq\psi\sq\varphi
+ \frac{1}{3}\sq(\nabla^c\psi\nabla_c\varphi) -
\frac{2}{3}\psi R\sq\varphi -\frac{2}{3}R\nabla^c\psi\nabla_c\varphi +
\frac{1}{3}\psi\nabla^cR\nabla_c\varphi \right.\right.\nonumber\\
&&
\qquad + 2R^{cd}\nabla_c\psi\nabla_d\varphi +
 2\psi R^{cd}\nabla_c\nabla_d\varphi\bigg) +
2\nabla_a\sq\psi\nabla_b\varphi +
2\nabla_a\sq\varphi\nabla_b\psi  -
2\nabla^c\psi \nabla_c\nabla_a\nabla_b\varphi -
2\nabla^c\varphi\nabla_c\nabla_a\nabla_b\psi \\
&&\left. +\frac{4}{3} \nabla_a\nabla_b(\nabla^c\psi \nabla_c\varphi )
- 2 \sq\psi \nabla_a\nabla_b\varphi - 2 \sq\varphi\nabla_a\nabla_b\psi  +
4 R_a^{\ c}(\nabla_c\psi \nabla_b\varphi +
\nabla_c\varphi\nabla_b\psi ) - \frac{4}{3}
R_{ab}\nabla^c\psi \nabla_c\varphi - \frac{4}{3} R
\nabla_a\psi \nabla_b\varphi \right]\nonumber\,,
\end{eqnarray}

\begin{eqnarray}
{\delta S_4^{(F)}} &=& \frac{b}{2} \int d^4x\,\sqrt{-g}\, h^{ab}
\left\{-2g_{ab}\left[\frac{1}{3}\sq^2\psi +
R^{cd}\nabla_c\nabla_d\psi  - \frac{2}{3}R\sq\psi  -
\frac{1}{6}\nabla^cR\nabla_c\psi \right] \right.\nonumber\\
&&
-2\left[\psi \left(2R^{cd}R_{acbd} - \frac{2}{3}RR_{ab} + \sq R_{ab}
- \frac{1}{3}\nabla_a\nabla_b R\right) + \frac{4}{3}R_{ab}\sq\psi
+ \frac{2}{3}R\nabla_a\nabla_b\psi \right. \nonumber\\
&& \left.\left. - 2R_a^{\ c}\nabla_b\nabla_c\psi + 2\nabla^c\psi
(\nabla_cR_{ab} - \nabla_a R_{bc}) -
\frac{1}{3}\nabla_a\nabla_b\sq\psi \right]\right\} \,.
\end{eqnarray}

{}From the last three equations one can read off the formula for
the $C_{ab}$ tensor, {\it i.e.}

\begin{eqnarray}
&& C_{ab} = -2R^{cd}C_{acbd}\psi
+4\nabla^c\nabla^d(C_{c(ab)d}\psi)
- \frac{1}{2}g_{ab}F_4\psi  +2g_{ab}\left(\frac{1}{3}\sq^2\psi
+ R^{cd}\nabla_c\nabla_d\psi  - \frac{2}{3}R\sq\psi
- \frac{1}{6}\nabla^cR\nabla_c\psi \right) \nonumber\\
&&
+ \psi \left(4R^{cd}R_{acbd} - \frac{4}{3}RR_{ab}
+ 2{\sq}R_{ab} - \frac{2}{3}\nabla_a\nabla_b R\right)
- \frac{2}{3}\nabla_a\nabla_b\sq\psi
+ \frac{8}{3}R_{ab}\sq\psi
+ \frac{4}{3}R\nabla_a\nabla_b\psi
- 4R^c\,_{(a}\nabla_{b)}\nabla_c\psi  \nonumber\\
&&
+ 4\nabla^c\psi (\nabla_cR_{ab} - \nabla_{(a}R_{b)c})
 + g_{ab}\biggl(\psi\sq^2\varphi  - \sq\psi\sq\varphi
+ \frac{1}{3}\sq(\nabla^c\psi\nabla_c\varphi)
- \frac{2}{3}\psi R\sq\varphi
-\frac{2}{3}R\nabla^c\psi\nabla_c\varphi
+ \frac{1}{3}\psi\nabla^cR\nabla_c\varphi \nonumber\\
&&
+ 2R^{cd}\nabla_c\psi\nabla_d\varphi
+ 2\psi R^{cd}\nabla_c\nabla_d\varphi\biggr)
- 2\nabla_{(a}\sq\psi\nabla_{b)}\varphi
- 2\nabla_{(a}\sq\varphi\nabla_{b)}\psi
+ 2\nabla^c\psi \nabla_c\nabla_a\nabla_b\varphi
+ 2\nabla^c\varphi\nabla_c\nabla_a\nabla_b\psi  \\
&&
+ 2 \sq\psi \nabla_a\nabla_b\varphi + 2 \sq\varphi\nabla_a\nabla_b\psi
-\frac{4}{3} \nabla_a\nabla_b(\nabla^c\psi \nabla_c\varphi )
- 4 R^c\,_{(a}(\nabla_c\psi \nabla_{b)}\varphi
+ \nabla_c\varphi\nabla_{b)}\psi )
+ \frac{4}{3} R_{ab}\nabla^c\psi \nabla_c\varphi
+ \frac{4}{3} R\nabla_{(a}\psi \nabla_{b)}\varphi \nonumber\,.
\end{eqnarray}
Using the definitions of $\varphi$ and $\psi$ one can verify that
the trace of the $C_{ab}$ tensor as given above is local and equal to
\begin{equation}
g^{ab}C_{ab} = C^{abcd}C_{abcd} = F_4\,,
\end{equation}
in agreement with (\ref{Ctrace}) of the text.

\begin{references}
\bibitem{Strom} As this paper was being submitted, the authors learned
that a similar conclusion has been reached by A. Strominger in hep-th/0106113.
\end{references}
\end{document}